\documentclass[bst/sn-mathphys-num]{sn-jnl}% Math and Physical Sciences Numbered Reference Style 
\usepackage{xcolor}%
\usepackage{soul}%
\usepackage{graphicx}%
\usepackage{multirow}%
\usepackage{amsmath,amssymb,amsfonts}%
\usepackage{amsthm}%
\usepackage{mathrsfs}%
\usepackage[title]{appendix}%
\usepackage{xcolor}%
\usepackage{textcomp}%
\usepackage{manyfoot}%
\usepackage{booktabs}%
\usepackage{algorithm}%
\usepackage{algorithmicx}%
\usepackage{algpseudocode}%
\usepackage{listings}%
\usepackage{subcaption}%
\usepackage{comment}
\theoremstyle{thmstyleone}%
\theoremstyle{thmstyletwo}%

\theoremstyle{thmstylethree}%

\begin{document}

\title[Article Title]{Infinite Scrolling, Finite Satisfaction: Exploring User Behavior and Satisfaction on Social Media in Bangladesh}

\iffalse
\author*[1,2]{\fnm{First} \sur{Author}}\email{iauthor@gmail.com}

\author[2,3]{\fnm{Second} \sur{Author}}\email{iiauthor@gmail.com}
\equalcont{These authors contributed equally to this work.}

\author[1,2]{\fnm{Third} \sur{Author}}\email{iiiauthor@gmail.com}
\equalcont{These authors contributed equally to this work.}

\affil*[1,2]{\orgdiv{Department}, \orgname{Organization}, \orgaddress{\street{Street}, \city{City}, \postcode{100190}, \state{State}, \country{Country}}}

\affil[2]{\orgdiv{Department}, \orgname{Organization}, \orgaddress{\street{Street}, \city{City}, \postcode{10587}, \state{State}, \country{Country}}}

\affil[3]{\orgdiv{Department}, \orgname{Organization}, \orgaddress{\street{Street}, \city{City}, \postcode{610101}, \state{State}, \country{Country}}}
\fi

\author*[2,1]{\fnm{Sanzana Karim} \sur{Lora}}\email{sanzana.lora@ewubd.edu}

\author[1]{\fnm{Sadia Afrin} \sur{Purba}}\email{sadia.afrin.purba@gmail.com}

\author[1]{\fnm{Bushra } \sur{Hossain}}\email{0412052016@grad.cse.buet.ac.bd}

\author[1]{\fnm{Tanjina } \sur{Oriana}}\email{tanjina.oriana@gmail.com}

\author[3]{\fnm{Ashek } \sur{Seum }}\email{seum.cse@aust.edu}

\author[1]{\fnm{Sadia } \sur{Sharmin}}\email{sadia@teacher.cse.buet.ac.bd}

\affil*[1]{\orgdiv{Department of Computer Science and Engineering}, \orgname{Bangladesh University of Engineering and Technology}, \orgaddress{ \city{Dhaka}, \country{Bangladesh}}}

\affil*[2]{\orgdiv{Department of Computer Science and Engineering}, \orgname{East West University}, \orgaddress{ \city{Dhaka}, \country{Bangladesh}}}

\affil[3]{\orgdiv{Department of Computer Science and Engineering}, \orgname{Ahsanullah University of Science and Technology}, \orgaddress{\city{Dhaka}, \country{Bangladesh}}}

\abstract{Social media platforms continue to change our digital relationships nowadays. Therefore, recognizing the complex
consequences of infinite scrolling is essential. This paper explores two distinct angles of social media engagement: mindless scrolling and mindful scrolling. This extensive study dives into numerous aspects of social media user behavior and satisfaction via the perspective of multiple surveys. We investigate
the psychological exploit of infinite scrolling design to keep users engaged, illuminating its effect on users’ emotional well-being. Furthermore, we explore its diverse effects on various groups,
such as teenagers, professional people, and pregnant women,
to better understand how digital activity differs throughout
life phases. Furthermore, our study reveals the psychological
consequences of being exposed to unfavorable news material. In
the context of nutritional objectives, we examine the problems
users confront as well as the significance of scrolling in dietary
achievement. By taking into account the demographic effect, we
can determine how factors like age, gender, and socioeconomic
position affect user behavior. This study presents a comprehensive
knowledge of the complicated connection of infinite scrolling with user satisfaction and psychological well-being through a variety of surveys, opening the door for well-informed conversations on
online engagement.}




\maketitle

\section{Introduction}\label{sec:intro}
Social media has ingrained itself firmly into modern life, influencing how people communicate, use information, and spend their time.  Currently, social media addiction is believed to affect 210 million individuals worldwide\footnote{\url{https://www.addictionhelp.com/social-media-addiction/statistics/}}.  Initially designed to facilitate communication and information sharing, social media platforms today function as part of the attention economy, which makes a profit off of user engagement.  Concerning this, platform designs have adjusted to take advantage of psychological processes that draw in and hold users' attention.

%Social media has taken center stage in the modern lifestyle, influencing all aspects of daily life and having a significant impact on how people interact and absorb information. As a result, an estimated 210 million individuals worldwide are addicted to social media\footnote{\url{https://www.addictionhelp.com/social-media-addiction/statistics/}}. Social media were created as web-based tools to promote messaging and information exchange among individuals. At the same time, many social media companies, such as Facebook, Twitter, and YouTube, rely on the continual attention of users to generate revenue, a phenomenon known as the attention economy. As a result, these technologies are now built to be inherently attractive to capture people’s attention.

Infinite scrolling, a design feature that constantly loads new content and bypasses natural stopping indications, is a prime illustration of this.  This method psychologically ties users to continuous action by taking advantage of the Zeigarnik Effect, which claims that individuals remember unfinished activities better.  In a similar vein, intermittent reinforcement theories and dopamine feedback loops explain how users become obsessive scrollers due to unpredictable and varying content rewards.  These tactics are part of attention engineering, which is the purposeful use of psychological concepts to influence user behavior and increase involvement.
%One of the most recognized design strategies is infinite scrolling, which offers consumers an unlimited stream of content. This feature encourages visitors to browse for long periods, often with no obvious destination. Endless scrolling makes use of psychological factors such as the Zeigarnik effect, in which people recall incomplete activities better than completed ones, and the dopamine feedback loop, which enhances the joy of discovering new content. These design decisions are crucial to what is known as "attention engineering," which involves manipulating psychological triggers to increase user engagement.

The important features of attention engineering encompass several factors. One is called Intermittent Reinforcement. By infrequently giving incentives, this strategy keeps users engaged by fostering suspense and uncertainty. People still browse in the hopes of finding worthwhile information, even though they don't always feel satisfied right away. Another one is called Personalization, which refers to the creation of content that is relevant to each user's interests. Social media companies employ algorithms that collect information on user activity, preferences, and demographics. Due to the great degree of customization, users spend more time on the site, since the content looks more relevant and interesting. The last is an addictive design. 

The design of social media platforms, which includes features such as effortless scrolling and auto-refreshing feeds, contributes to long-term engagement. These design aspects reduce the effort necessary to obtain new content, while also promoting a fear of missing out (FOMO), encouraging users to stay connected constantly.

This paper draws on these theoretical foundations to investigate two different kinds of scrolling behavior to investigate the implicit consequences of these design features on user behavior and satisfaction in Bangladesh:
\begin{itemize}
    \item \textbf{Mindless Scrolling: }Specified by unaware, frequently obsessive browsing powered by addictive design aspects.
    \item \textbf{Mindful Scrolling: }Consists of deliberate, purposeful engagement with content that is relevant to the user's interests and goals.
\end{itemize}

The contribution of our paper is as follows.
\begin{itemize}
    \item \textbf{Observation and Analysis:} Examines the ways that design features like endless scrolling affect user behavior and psychological well-being in Bangladesh.
    \item \textbf{Behavioral Comparisons:} Illustrates when scrolling is mindless or mindful, and offers insights on how each affects user satisfaction and mental well-being in Bangladesh.
    \item \textbf{Demographic Insights:} Examines the experiences and reactions of several demographic groups, including professionals, pregnant women, and teenagers, to diverse scrolling patterns in Bangladesh.
    \item \textbf{Design Implications:} Offers suggestions on developing social media features that encourage more positive interaction and minimize detrimental psychological impacts.
\end{itemize}

\section{Related Works}\label{sec:related}

The addictive design of social media platforms hacks the human soul and leads to infinite scrolling. A study found that more than a billion people spent an average of three hours on social media scrolling in the year 2020 \cite{purohit2021unhooked}. Users are not self-aware while technology controls them. Many researchers analyzed different social media design patterns, user satisfaction, and the positive and negative effects of it on user behavior.

Roffarello and Russis identified five dark patterns, such as suggestions, autoplay, pull-to-refresh, infinite scrolling, and social investment, that exploit psychological vulnerabilities to increase the time spent on digital services \cite{monge2022towards}.

Caraban et al. analyzed ways of manipulating and biasing decision-making using technology nudging. Positioning, coloring, opt-out features, information hiding like clickbait, suggesting multiple options, friction creation, keeping minor alternatives so that users are more likely to choose the targeted option, creating many viewpoints, deceiving, provoking decisions made on previous experiences, providing features where people can compare themselves, use of uncertainty and reminding which influence to scroll mindlessly \cite{caraban201923}. 

Wang indicated that scrolling up features while streaming videos trigger psychological patterns in users and they end up in a loop. Short videos and reels are included in the design as it involves more users scrolling continuously \cite{article}. 

Adams et al. explained the influence of persuasive technology on humans and explained that 'Automatic mind' and 'reflective mind' control human behavior. Different technologies were introduced to motivate users by showing good and positive things in the news feed. They suggested reflexive designs which can keep the focus, have some triggering elements, parallel functionality for attracting users, and impact on their behavior \cite{adams2015mindless}.

Recently, several studies demonstrated the impact and influence of using social media on different social media users by conducting several surveys.

Kumar et al. examined the impact of augmented reality (AR)-based filters on social media users’ self-concept and well-being by utilizing the inductive qualitative method and grounded theory to analyze 18 AR filter users’ in-depth interviews \mbox{\cite{filtering_revref1}}.

Pang et al. developed a conceptual study model that focuses on how two unique aspects of interactivity - social interactivity and system interactivity - influence users' perceived advantages from mobile short video apps, eventually influencing their desire to continue using them. This investigation thoroughly examined data collected from 808 users of mobile short video applications, using structural equation modeling to support the proposed hypothesis \mbox{\cite{PANG2024103923_revfer2}}.

In another study, Pang et al. meticulously dissected the implications of overall service quality on WeChat user assessments, encompassing aspects such as user satisfaction, identification, belongingness, and perceived benefit, while identifying the correlation between user assessments and their proclivity towards emotional and service attachment. The research paradigm was carefully verified using data from the Sojump database, which includes 598 reliable observational data points.  The study model was then confirmed and certified using the robust structural equation modeling approach \mbox{\cite{PANG2024103688_revref3}.}

To thoroughly assess the effects of cumulative satisfaction and stickiness on electronic word-of-mouth engagement (eWOM), Pang et al. utilized motivation theory and the expectation disconfirmation approach to systematically examine the dynamic influences from multidimensional perceived benefits for example, functional benefits, psychosocial benefits, and hedonic benefits on WeChat users' cumulative satisfaction on 689 active WeChat users in mainland China \mbox{\cite{PANG2024102174_rvwref4}}.

The fundamental relationships between youth civic engagement, network heterogeneity, network capital, and mobile social media use were systematically investigated by Pang et al.  1,208 youths in mainland China participated in a web-based survey that yielded the findings \mbox{\cite{pang_revref5}}.

In a cross-sectional study including 898 students at universities, Pang et al. discovered that self-disclosure and depression exacerbated social and information overload, which in turn caused problematic mobile app use and poor academic performance.  The association between self-disclosure and excessive app use was additionally influenced by social overload \mbox{\cite{pang2024determining_revref6}}.

Pang et al. pioneered the use of the SSO framework to research technological stress in university students during mobile-assisted learning, addressing a significant gap in prior research that mostly focused on corporate or academic populations.  The conceptual model has been experimentally confirmed using a comprehensive study of statistics from 605 university students in mainland China \mbox{\cite{PANG2025102861_revref7}}.

Furthermore, some researchers were concerned about how these media platforms can be scrolled mindfully. Purohit and Holzer designed an ethical nudging interference that can minimize the time spent on social media and encourage mindfulness practice \cite{purohit2020designing}. Their survey strongly indicates that it can develop positive habits and encourage users to reduce the overuse of social platforms. In another work, Purohit et al. developed a browser extension that made social media less compelling \cite{purohit2021unhooked}. News feed diets can help users stop zombie scrolling \cite{purohit2023starving}.

Previous studies show some design mechanisms that inspire to scroll mindless scrolling and some interventions in nudges that discourage infinite scrolling. In most cases, the survey was conducted on a specific group of people, which may not provide a clear concept of various perspectives. In this research, the focus is on analyzing information in the local context. %and compare it with the global views. 
Furthermore, how both mindful and mindless practices have impacts on people's behavior needs to be analyzed at the same time and on the same user. The survey should include users from different groups. The research was conducted including these parameters.

\section{Our Study: Mindless and Mindful Scrolling of Social Media}
\label{study}
In this section, we thoroughly discuss our study on mindless and mindful scrolling on social media. We accomplish our study considering several factors like anxiety and fatigue levels, the psychological well-being of teenagers, demographic factors, professionals,  women's views on the pregnancy period, food trends, and the design principles and features of social media to facilitate digital detoxification. We divide them into several research questions to explore user behavior and satisfaction on social media. 

\subsection{\textbf{Study on Mindless Scrolling}}
This study addresses the following research questions to comprehend the effects of mindless social media scrolling. Survey results from a variety of demographic groups were used to get the answers explained in Section \ref{outcome mindless}.

\subsubsection{\textbf{RQ1: Does the nature of the content encountered during infinite scrolling, such as negative news, triggering posts, etc., impact users' anxiety and fatigue levels? }}
\label{rq1}

Users' feelings of anxiety and fatigue can be affected by the nature of the content they encounter while using infinite scrolling. Several studies and observations have demonstrated that users' continual exposure to negative news, triggering posts, and other information on social media platforms and websites can boost their feelings of anxiety, stress, and mental fatigue\cite{anto2023exploring}\cite{BRIGHT2015148}. This study aims to determine whether the sort of content users encounter when infinite scrolling, such as negative news, upsetting postings, and other kinds of content, has any detectable effects on their levels of anxiety and fatigue. Many platforms' algorithms frequently tailor content to meet users' current thoughts and tastes. This can result in confirmation bias, in which people are exposed exclusively to content that supports their beliefs. While this may offer a sense of approval, it can also lead to polarized views and increased anxiety when faced with contradictory content.
\paragraph{\textbf{Method}}
\begin{itemize}
    %\item \textbf{Survey Instrument:} To collect data, we created an online survey with Google Forms ensuring accessibility to a diverse range of social media users.
    \item \textbf{Survey Design: }An organized online survey was created with Google Forms to determine the emotional and psychological impact of information received while infinite scrolling.  The instrument comprised four sections:

 \begin{itemize}
     \item Demographics: include age, gender, education, and employment.
     \item Social Media Usage: includes time spent scrolling, platforms utilized, and frequency of involvement.
     \item Content Nature: The types of contents that are regularly viewed (for example, unfavorable news, personal incidents), as well as users' subjective perceptions of content tone (positive, neutral, or negative).
     \item Emotional Outcomes: Self-reported levels of anxiety and fatigue were assessed %using a 5-point Likert scale 
     resulting from normal scrolling sessions.
 \end{itemize}

%\hl{ Questions were derived from current literature on digital media and psychological well-being, then pretested with a small group to ensure clarity and relevance.}

\item \textbf{Sampling and Data Collection:} A convenient sampling strategy was utilized to target ordinary adult social media users in Bangladesh. The survey was circulated by email, social media, and community networks.
\begin{itemize}
    \item Sample Size: Responses were obtained from
67 social media users from Bangladesh.
\item Participants were ensured of anonymity and provided informed consent.
\end{itemize}
%\item \textbf{Sample Size:} Convenience sampling was used, and responses were obtained from 67 social media users from Bangladesh.
%\item \textbf{Data Collection:} Participants' infinitely scroll timelines' and the types of content they frequently encounter, and their emotional responses to this content are asked in this survey.
\item \textbf{Data Analysis:} Data analysis included the use of descriptive statistics to summarize demographic information and investigate possible associations between infinite scrolling and the emotional responses of users.
\end{itemize}

\subsubsection{\textbf{RQ2: How does the infinite scrolling feature influence the online behavior, engagement pattern, and psychological well-being of teenage users of social media platforms?
}}
\label{rq2}
Digital media design addictive technology that exploits psychological vulnerabilities to hold attention and maximize screen time for two purposes. One is to push advertise and sell, and another is to extract user data for sale. The infinite scrolling feature of social media platforms significantly influences the online behavior, engagement patterns, and psychological well-being of teenage users. The platforms' algorithm-driven content delivery, exposure to a wide array of unfamiliar teenagers, and the ensuing social comparison dynamics contribute to positive and negative effects on teenagers. Moreover, young people have psychological vulnerabilities. For example, evolutionary biases dictate that we should accomplish what is simple and enjoyable rather than what is difficult and tedious. Social pressure - all of their friends are online. Young, developing brains are readily persuaded and lack self-control. Addiction-prone neurological disorders. This study investigates the influence of the infinite scrolling feature on teenagers.

\paragraph{\textbf{Method}}

\begin{itemize}
\item \textbf{Survey Design: } Considering the limited availability of online surveys in institutional settings, a specially designed paper-based questionnaire was developed for teenager respondents. There were 21 questions in the survey covering the following dimensions:
\begin{itemize}
    \item Demographics include age, gender, and academic level.
    \item Usage patterns include daily screen time, favorite platforms, and scrolling tendencies.
    \item Behavioral and emotional effects include changes in attention span, sleep pattern, self-control, engagement types (passive vs. active), and psychological indications such as loneliness, social comparison, or stress.
    \item The questions were age-appropriate and approved by educational institutions and instructors for sensitivity and readability.
\end{itemize}
    %\item \textbf{Survey Instrument:} We conducted our survey of school students between the ages of 15 to 21 in urban areas of Bangladesh. We took permission from their respective authorities. We offered enthusiastic students to participate in a pen and paper-based survey containing 21 questions. 
    %\item \textbf{Sample Size:} Responses were obtained from 61 teenagers.
    \item \textbf{Sampling and Data Collection:} The survey was undertaken among urban Bangladeshi school and college students aged 15 to 21. Before conducting the survey, permission was acquired from the institutions. Participation was entirely voluntary.
    \begin{itemize}
        \item Sample Size: The survey was filled out by 61 teenagers.
        \item The distribution and collection of surveys took place under supervision during school hours.
    \end{itemize}
    \item \textbf{Data Analysis:} Data analysis included descriptive statistics to summarize demographic information and investigate possible associations between infinite scrolling and the emotional responses of teenagers.
    
\end{itemize}
\begin{comment}

\subsubsection{\textbf{RQ3: How do different demographic factors influence users' engagement patterns and responses to infinite scrolling in social media?}}
\label{rq3}
Different users from different demographics can have
variations in engagement patterns \cite{demo}. An old man and young boy
may not find interest in the same content or in the same design
pattern which is designed to draw attention. Gender and Economic status have also impacts on social media scrolling patterns.
\paragraph{\textbf{Methodology}}
\begin{itemize}
    \item \textbf{Survey Instrument:} Data was collected considering different ages, gender, and economic status. We created an online survey with Google Forms ensuring accessibility to a diverse range of social media users.
    \item \textbf{Sample Size:} Convenience sampling was used, and responses were obtained from 67 social media users in Bangladesh.
    \item \textbf{Data Collection:} This survey included a questionnaire about how demographic factors influence user behavior and impact engagement. 
    \item \textbf{Data Analysis:} Data analysis included the use of descriptive statistics to summarize demographic information and how users think about the demographic influence on using social media.
\end{itemize}
\end{comment}

\subsubsection{\textbf{RQ3: How does implementing infinite scrolling on social media platforms impact the productivity, information consumption habits, and overall work performance of professional individuals within a workplace context?
}}
\label{rq4}

We all want to make the best use of our time, particularly at work. Despite this, our studies reveal that 80\% of Bangladeshi respondents who are working in different professions use social media while at work, with many using it for many hours every day. Cyberloafing refers to employees who use work time to participate in non-work-related online activities such as web browsing, social media use, or monitoring and replying to personal emails. According to University of Nevada researchers \cite{wsj}, it costs organizations an estimated $85$ billion per year. According to a 2018 analysis by Udemy, an online learning company, over 62\% of survey participants \cite{udemy} spend about an hour staring at their phones during the workday. 

This study investigates the influence of the infinite scrolling feature of social media platforms in the workplace. 

\paragraph{\textbf{Method}}

\begin{itemize}
    %\item \textbf{Survey Instrument:} To collect responses from professionals, we created an online survey with Google Forms ensuring accessibility to a diverse range of social media users.
    \item \textbf{Survey Design: }
 An online survey was distributed to Bangladeshi professionals working in several fields.  The instrument focuses on:
 \begin{itemize}
     \item Demographics: Age, occupation, and industry.
     \item Workplace Social Media Use: The frequency and timing of social media use during business hours.
     \item Content Consumption Habits: The types of content seen, as well as the amount of time spent infinite scrolling during office hours.
     \item Productivity Impact: An assessment of job interruptions, task switching, and time wasted due to scrolling.
 \end{itemize}

 \item \textbf{Sampling and Data Collection: }Participants were approached via different professional networks such as LinkedIn, WhatsApp, Facebook groups, and emails.  The survey accentuated workplace confidentiality and voluntary participation.

\begin{itemize}
    \item Sample Size: 36 professionals from various sectors (e.g., IT, education, marketing) conducted the survey.
\end{itemize}
    %\item \textbf{Sample Size:} Responses were collected from 36 professionals working in different sectors in Bangladesh.
    \item \textbf{Data Analysis:} Data analysis included the use of descriptive statistics to understand the influence of social media on productivity.

\end{itemize}

\subsection{Study on Mindful Scrolling}
Mindful scrolling, which refers to the deliberate and attentive usage of social media and online content, can influence a person's emotional state and thinking. In this section, we study two research questions on mindful scrolling. The purpose of this study is to better understand the ramifications of mindful scrolling on social media by answering the research questions. The outcomes of this study are elaborated in Section \ref{outcome mindful}.

\subsubsection{{\textbf{RQ4: Does frequent mindful scrolling during pregnancy influence women's views of their emotional preparedness for parenting?}}}
\label{rq5}
Being a parent is a significant life event that is frequently accompanied by feelings of joy and apprehension. Individuals have access to many types of information and social media material in the era of digital technology, which can influence their opinions about motherhood. Several studies and observations have shown that social networks play a vital role during the pregnancy period \cite{al2021pregnancy}\cite{smith2020relationship}. This study investigates the possible impact of mindful scrolling practices on women's perceptions of their emotional preparation for parenthood during pregnancy.

\paragraph{\textbf{Method}}
\begin{itemize}
    \item \textbf{Survey Design: } We used Google Forms to create a structured online survey that was accessible to a diverse group of social media users.  The questionnaire was organized into the following sections:

    \begin{itemize}
        \item Demographics: Age, educational background, and if the respondent is presently pregnant or has been pregnant in the recent past.
        \item Social Media Usage: Platforms used, daily time spent on social media, and types of content consumed (for example, parenting advice, health updates).
        \item Scrolling Behavior: The frequency, duration, and particularly during emotionally vulnerable times.
        \item Emotional Impact: Questions about perceived emotional readiness for parenthood throughout pregnancy, stress levels, emotional exhaustion, and the role of digital content.
    \end{itemize}
 
    \item Sampling and Data Collection:  We utilized convenience sampling to reach respondents in Bangladesh through online parenting networks, social networks, and health forums.
    \begin{itemize}
        \item Sample Size: Valid responses were gathered from 36 participants, including both pregnant women and those who have previously been pregnant.
        \item All respondents provided informed consent and participated voluntarily.  No personally identifiable information was collected.
        
    \end{itemize}
 
    %\item \textbf{Survey Instrument:} To collect data from pregnant women, we created an online survey with Google Forms ensuring accessibility to a diverse range of social media users.
    %\item \textbf{Sample Size:} Convenience sampling was used, and responses were obtained from 36 pregnant women of Bangladesh.
    %\item \textbf{Data Collection:} Participants' scrolling patterns during pregnancy and opinions on their emotional preparedness for parenthood were questioned.
    \item \textbf{Data Analysis:} Data analysis included using descriptive statistics to summarize demographic information and investigate possible associations between mindful scrolling and emotional preparation during pregnancy.
\end{itemize}

\subsubsection{\textbf{RQ5: Does frequent mindful scrolling of social media food trend posts influence users to achieve a dietary goal?}}
\label{rq6}
In the digital era, social media platforms have a tremendous impact on users' food choices and goal-setting habits. Several studies and observations have demonstrated the impact of social media on food habit \cite{leu2022food}\cite{chung2017food}. While scrolling through food trends can inspire healthier habits and nutritional goal accomplishment, there are concerns about potential negative consequences. The aesthetically appealing content frequently highlights aesthetics over nutrition, potentially magnifying interest in unhealthy food choices. The convenience of ordering restaurant meals through these platforms may lead to fewer nutritional choices, and appealing food trends can provide significant obstacles for people trying to lose weight. This study dives into the complex interaction between thoughtful scrolling through food trends on social media and dietary patterns, intending to understand both good motives and possibly negative repercussions on current diet choices. 
\paragraph{\textbf{Method}}
\begin{itemize}
    \item \textbf{Survey Design: } To gather relevant data, we designed a survey online using Google Forms, making it accessible to a wide spectrum of social media users in Bangladesh.  The survey was organized into the following sections: 
    \begin{itemize}
        \item Demographics: Age, gender, dietary preferences, and so on.
        \item Social Media Behavior: Platforms preferred,  screen time, and frequency of engaging with food-related content.
        \item Food Trend Attention: Varieties of food trend content viewed (e.g., roadside food, diet trends, recipe reels), scrolling frequency, and level of consequence on personal food preferences.
        \item Dietary Goals: Questions that estimated whether participants set dietary plans, got inspiration after viewing food trends, and self-reported success in adhering to these goals.
        \item Negative Effects: Participants were also questioned to deliberate on examples in which social media scrolling led to poor diet decisions or disruptions in their goal-setting approach.
    \end{itemize}
    %\item \textbf{Survey Instrument:} To collect data, we created an online survey with Google Forms, ensuring accessibility to a diverse range of social media users.
    %\item \textbf{Sample Size:} Convenience sampling was used, and responses were obtained from 50 participants of varying ages and dietary preferences in Bangladesh.
    %\item \textbf{Data Collection:} The survey included questions about participants' social media activities, frequency of scrolling through food trend posts, dietary goal-setting behaviors, self-reported dietary goal successes, and any negative impacts on dietary choices.
    \item \textbf{Sampling and Data Collection: } We utilized convenience sampling to distribute the survey across several online platforms, particularly lifestyle communities and forums.
    \begin{itemize}
        \item A total of 50 participants met the survey, illustrating a mixture of age groups and dietary preferences from various areas of Bangladesh.
        \item Participation was unforced, and declared consent was received from all respondents.
    \end{itemize}
    \item \textbf{Data Analysis:} Data analysis included using descriptive statistics to summarize demographic information and investigate possible associations between mindful scrolling, dietary goal setting, and achievement.
\end{itemize}

\section{Outcomes of Our Study}
\label{outcome}
This section highlights the survey findings on mindless and mindful scrolling behaviors among social media users in Bangladesh as described in Section \ref{study}. The data gathered provides useful insights into how these opposing habits develop in various demographic groups, as well as their effects on users' emotional well-being and productivity. The following analysis goes into the prevalence, causes, and impacts of these habits, offering a thorough knowledge of how they influence users' digital lives.

\subsection{Outcomes of Mindless Scrolling}
\label{outcome mindless}
\subsubsection{Findings from RQ1}
\begin{figure}
     \centering
     \begin{subfigure}[b]{0.45\textwidth}
         \centering
         \includegraphics[width=\textwidth]{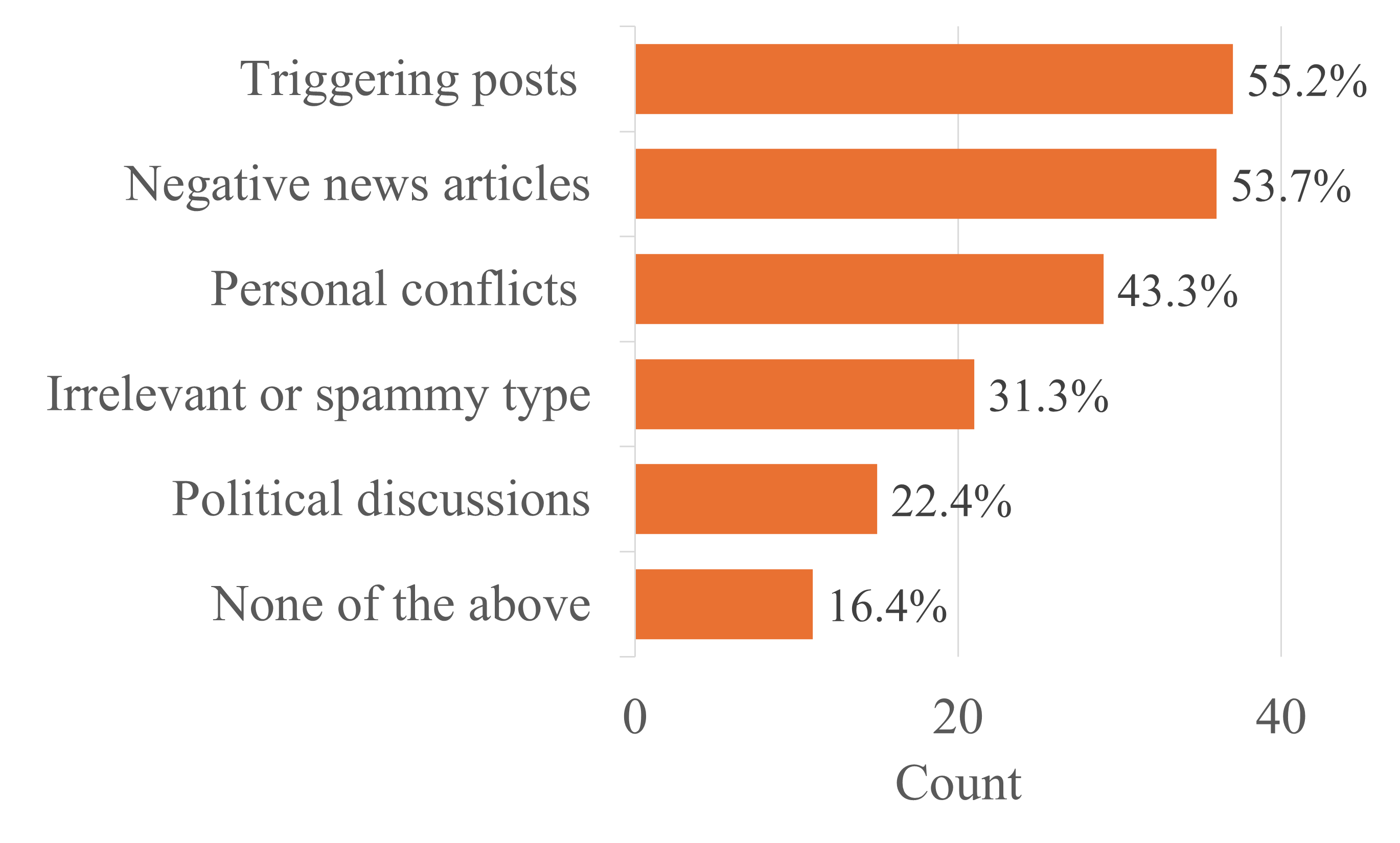}
         \caption{Types of content encountered during infinite scrolling which has the most significant impact on anxiety and fatigue levels}
         \label{fig:n1}
     \end{subfigure}
     \hfill
     %\hfill
     %\vspace{-1cm}
      \vspace{-0.5cm}
     \begin{subfigure}[b]{0.45\textwidth}
         \centering
         \includegraphics[width=\textwidth]{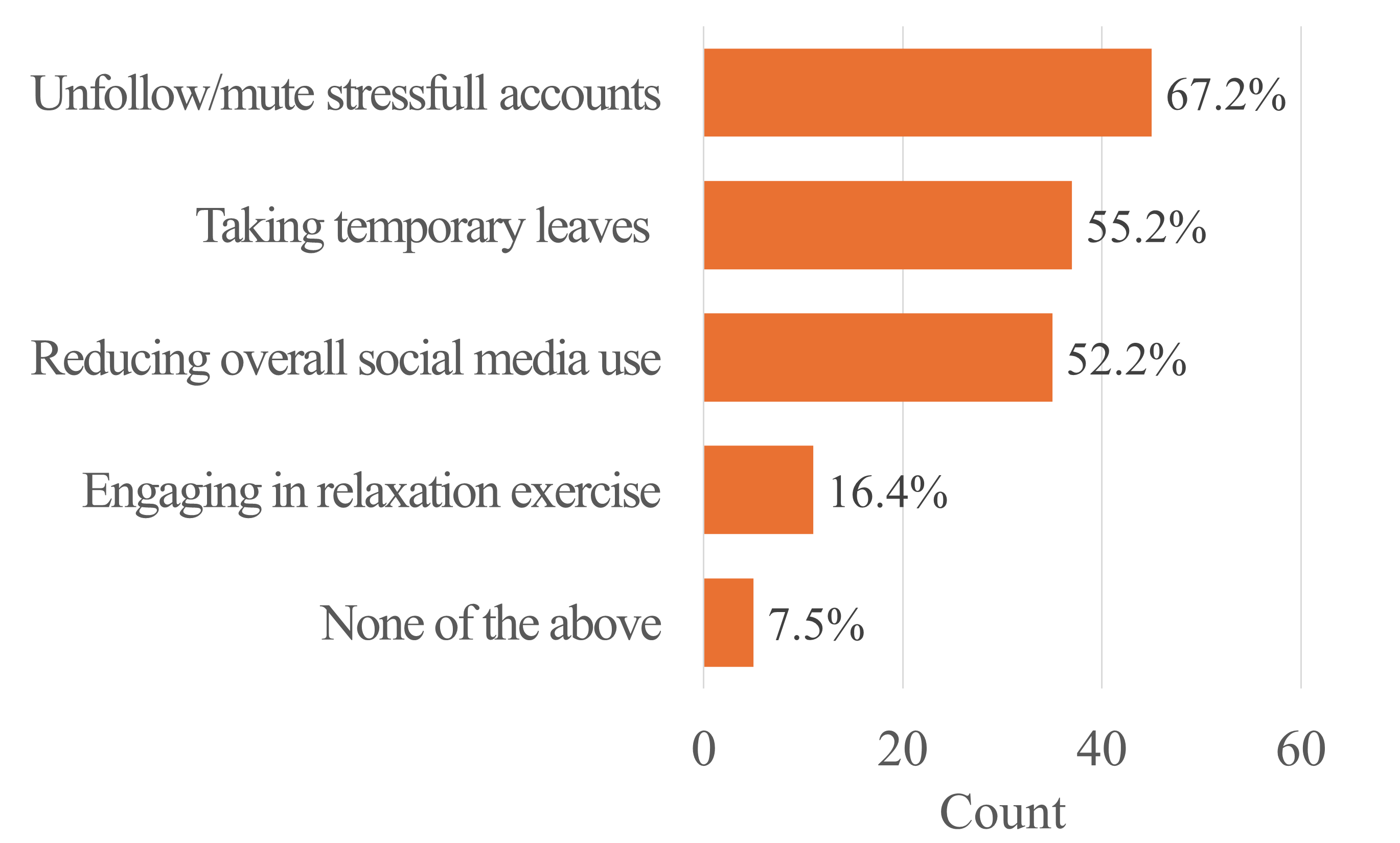}
         \caption{Strategies used to manage anxiety and fatigue resulting from social media infinite scrolling}
         \label{fig:n2}
     \end{subfigure}
     \hfill
     \vspace{0.5cm}
     \begin{subfigure}[b]{0.45\textwidth}
         \centering
         \includegraphics[width=\textwidth]{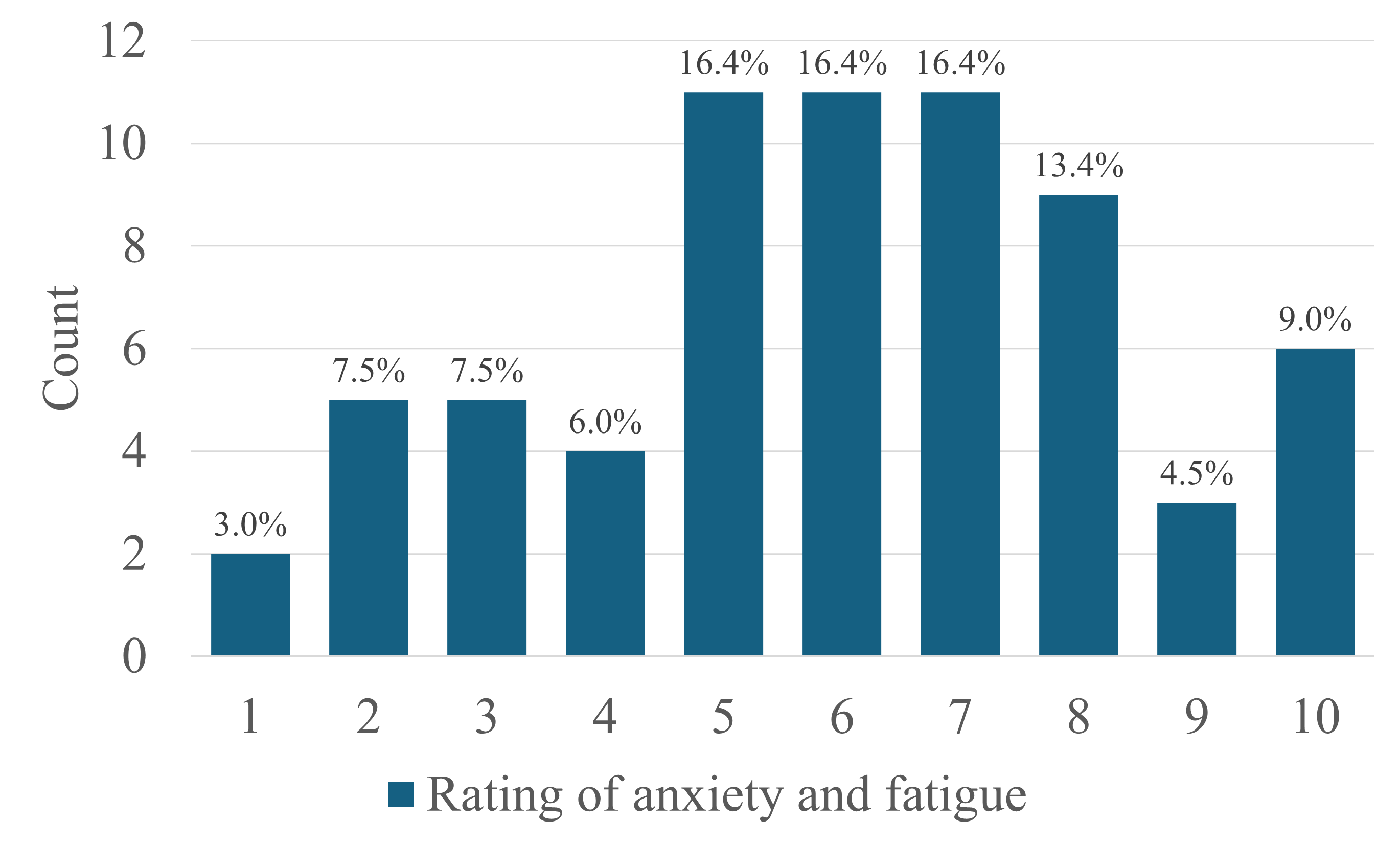}
         \caption{Rating of typical level of anxiety and fatigue after engaging stressful content in infinite scrolling on social media}
         \label{fig:n3}
     \end{subfigure}
     \hfill
     \hfill
     %\vspace{-0.5cm}
     \begin{subfigure}[b]{0.45\textwidth}
         \centering
         \includegraphics[width=\textwidth]{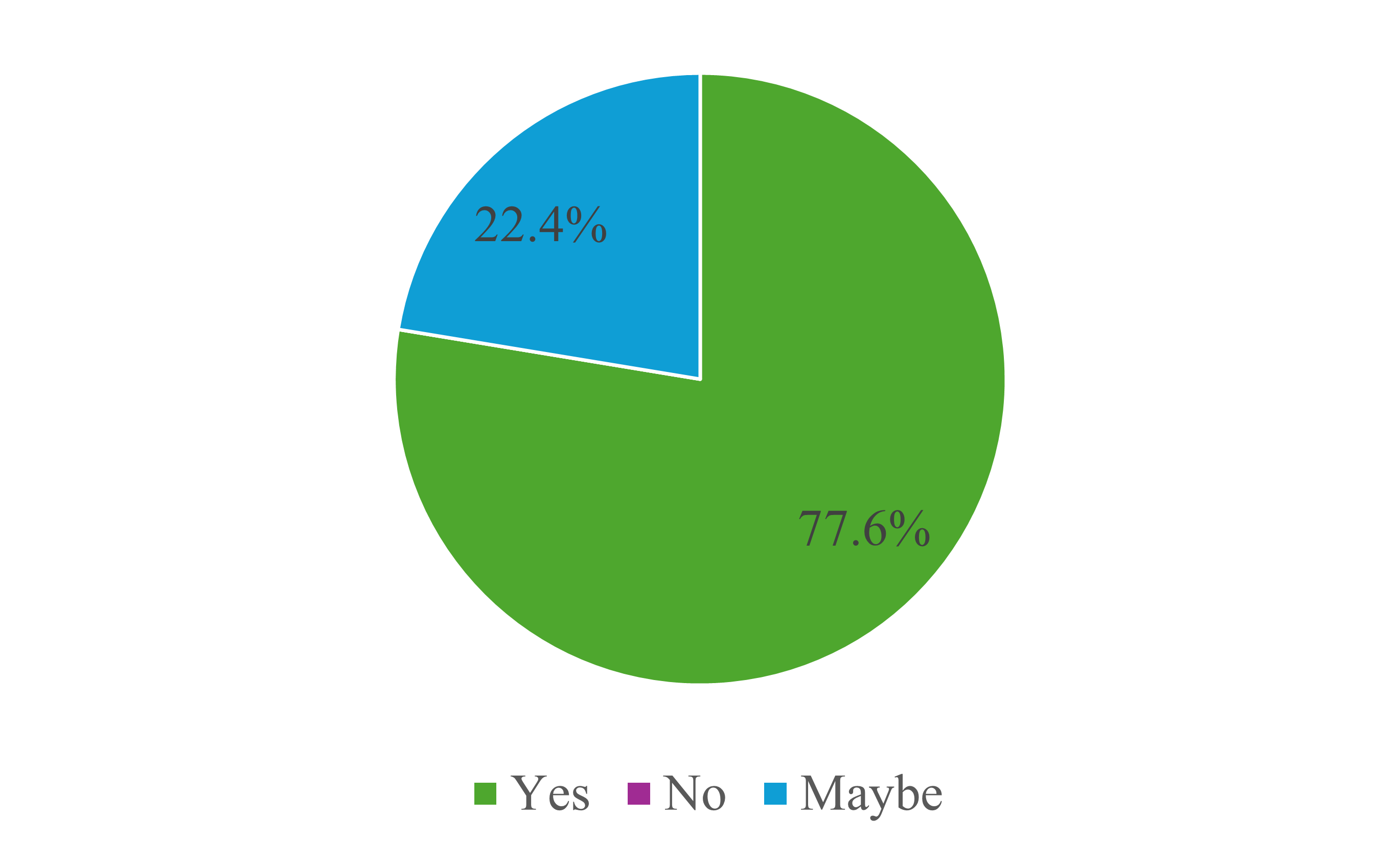}
         \caption{Social media platforms should provide tools to allow users or control the types of content they encounter during infinite scrolling to reduce anxiety and fatigue}
         \label{fig:n4}
     \end{subfigure}
        \caption{Our survey responses on the impact of infinite scrolling on anxiety and fatigue level of the nature of the content}
        \label{fig:neg}
\end{figure}

This section focuses on the key outcomes for RQ1, analyzing the nature of the content encountered during infinite scrolling, such as negative news, triggering posts, etc., impacting users’ anxiety and fatigue levels mentioned in Section \ref{rq1}. Figure \ref{fig:neg} refers to some significant responses of our study mentioned in Section \ref{rq1}.

\begin{itemize}
    \item \textbf{Participants' Demographics:} Respondents were mostly between the ages of 18 and 24 (74.6\%). 62.7\% of the respondents are male and the rest(33.3\%) are female.
    \item \textbf{Anxiety and Fatigue Level after Engaging Stressful Content:} On a scale of 1 to 10, with 1 being "not at all" and 10 being "extremely," the respondents were asked to rate their typical level of anxiety and fatigue after engaging stressful and negative content in infinite scrolling on social media. 16.4\% of respondents rated for 5, 16.4\% of were for 6, 16.4\% of were for 7, and 13.4\% were rated for 8 which indicates the increasing fatigue and anxiety level of users for infinite scrolling of stressful content.
    \item \textbf{Negative Emotional Impact: } Repeated exposure to unpleasant news items, distressing images, or triggering news can cause significant negative emotions. As users receive and absorb unpleasant information, they can experience increased anxiety, sadness, or even anger. 55.2\% of respondents reported that they have a significant impact on their anxiety and fatigue levels due to triggering posts. The proportion is 53.7\% for negative news articles. 
    %\item \textbf{Social Comparison: }Social media frequently portrays controlled representations of people's lives, highlighting mainly their good events. This can cause users to compare themselves unfavorably to others and have the feeling of missing out. This continual social comparison can add to feelings of insufficiency and anxiety.
    \item \textbf{Strategies to Manage Anxiety and Fatigue Level: } 67.2\% of our respondents followed the strategy to mute or unfollow the stressful content. 55.2\% of respondents took a break from social media to reduce anxiety and fatigue due to infinite scrolling. 77.6\% of respondents believed that social media platforms should provide tools or options to allow users to filter or control the types of content they encounter during infinite scrolling to reduce anxiety and fatigue levels.
\end{itemize}

\subsubsection{Findings from RQ2}
\begin{figure*}
     \centering
     \begin{subfigure}[b]{0.45\textwidth}
         \centering
         \includegraphics[width=\textwidth]{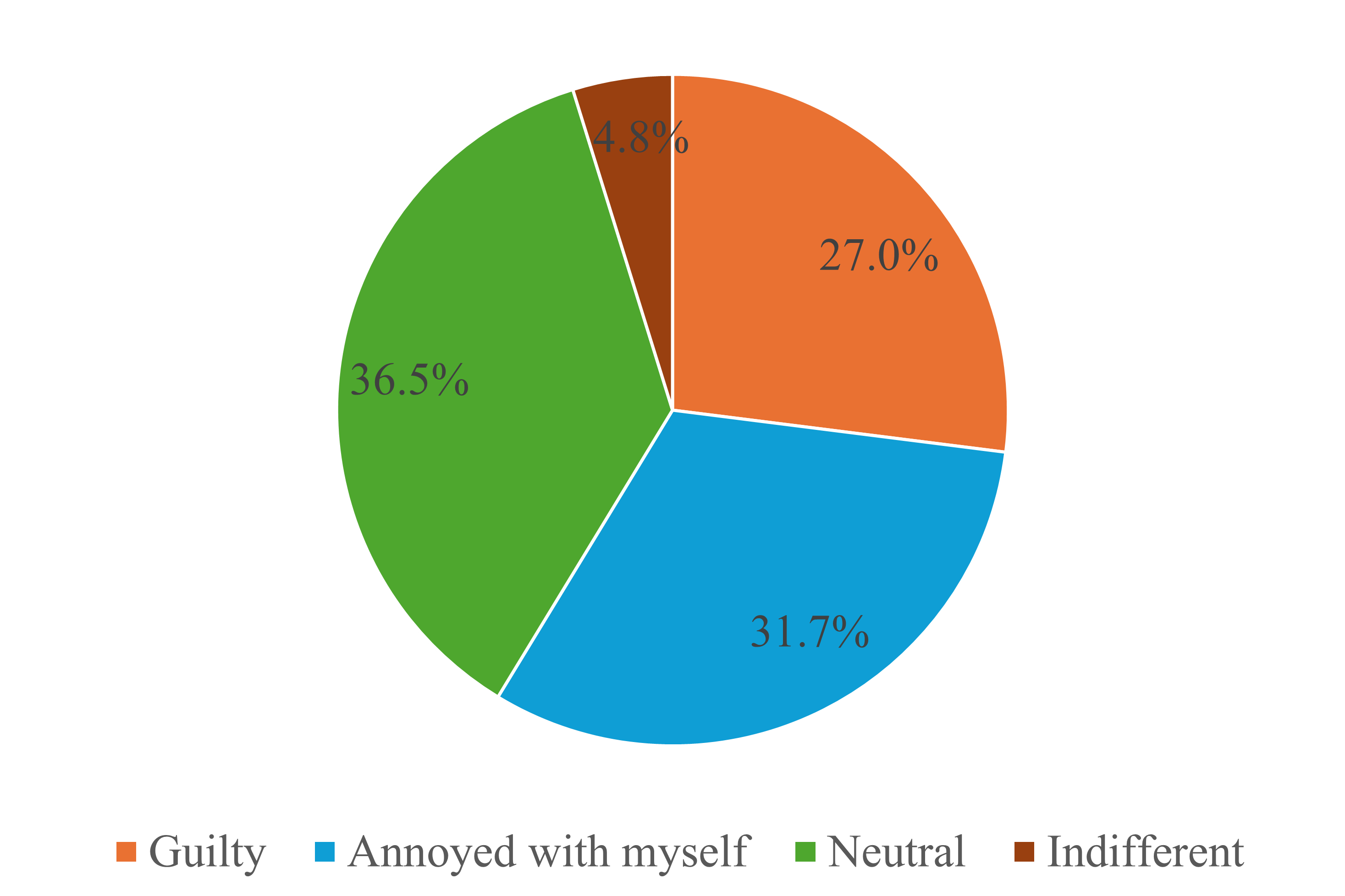}
         \caption{Types of emotions after mindless scrolling}
         \label{fig:teen-pie-1}
     \end{subfigure}
     \hfill
     \begin{subfigure}[b]{0.45\textwidth}
         \centering
         \includegraphics[width=\textwidth]{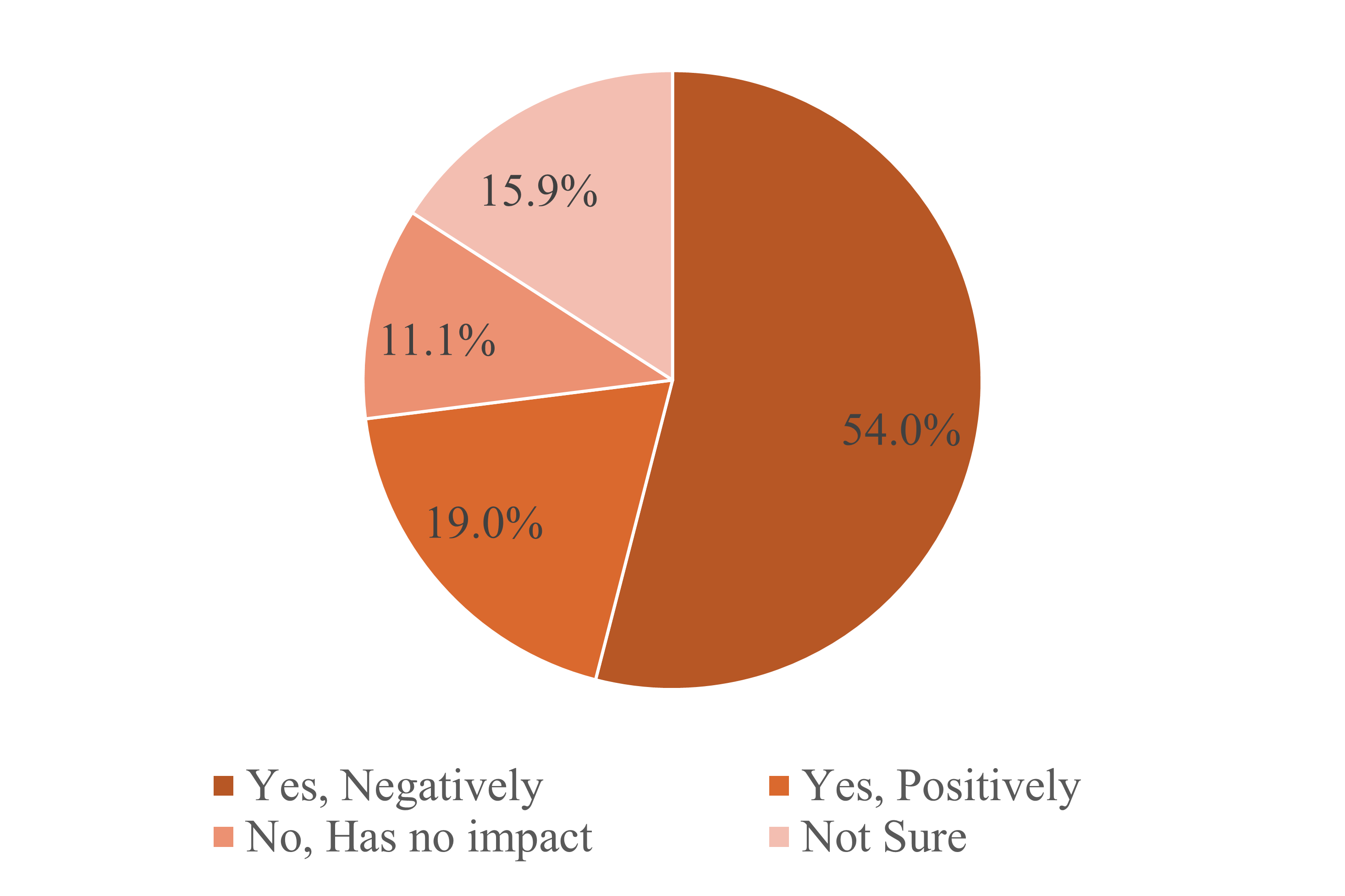}
         \caption{Response on their overall well-being after mindless scrolling}
         \label{fig:teen-pie-2}
     \end{subfigure}
     \hfill
     \vspace{0.5cm}
     \begin{subfigure}[b]{0.45\textwidth}
         \centering
         \includegraphics[width=\textwidth]{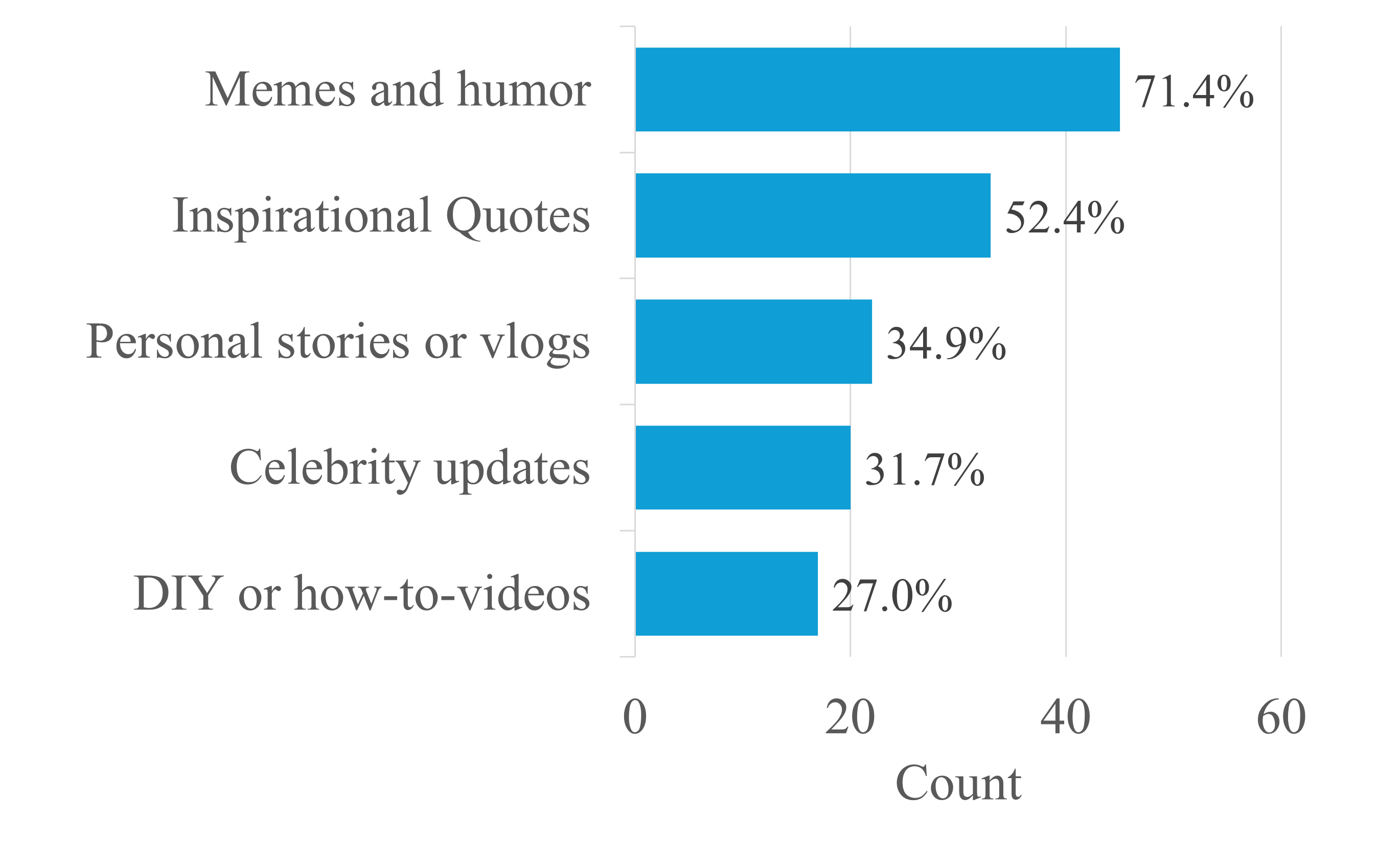}
         \caption{Types of content trigger mindless scrolling for teenage users}
         \label{fig:teen-bar-1}
     \end{subfigure}
     \hfill
     \begin{subfigure}[b]{0.45\textwidth}
         \centering
         \includegraphics[width=\textwidth]{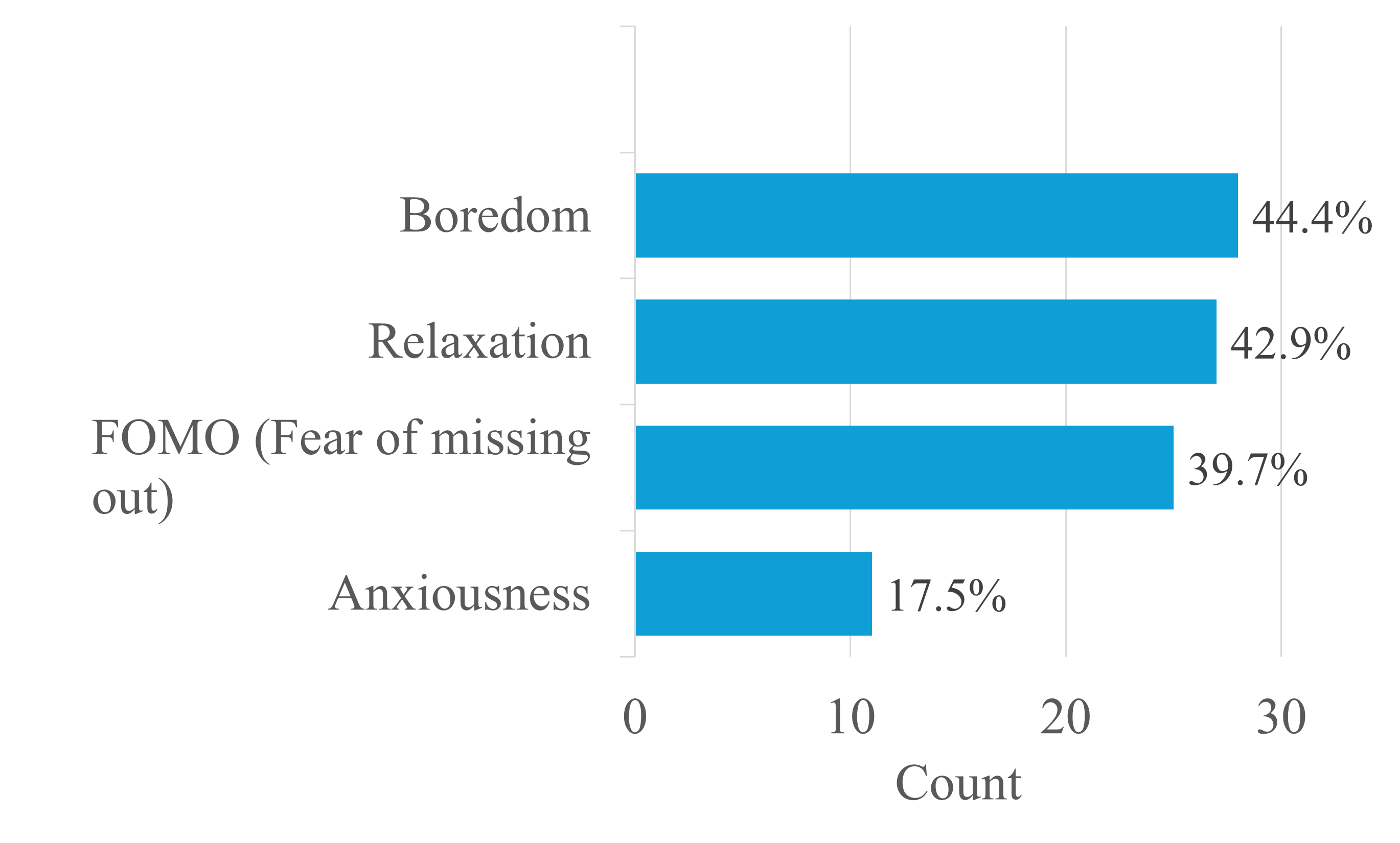}
         \caption{Types of emotions or feelings that trigger mindless scrolling}
         \label{fig:teen-bar-2}
     \end{subfigure}
        \caption{Some survey responses on the influence of infinite scrolling of social media on teenage users in Bangladesh}
        \label{fig:teenage}
\end{figure*}

This section focuses on the key outcomes for RQ2 analyzing the infinite scrolling feature influences the
online behavior, engagement pattern, and psychological
well-being of teenage users of social media platforms mentioned in Section \ref{rq2}. Figure \ref{fig:teenage} refers to some significant responses of our study mentioned in Section \ref{rq2}.

\begin{itemize}
    \item \textbf{Participants' Demographics:} The respondents belong to three age groups: 13-15 (59\%), 16-18 (19.4\%), and 19-21 (24.6\%), including both male 59\% and female (41\%) from \textbf{Bangladesh}.
    \item \textbf{Usage of Social Media: } 
    The study found that 100\% of participants use some form of social media platform such as Facebook, Instagram, YouTube, etc. Among these users, 68.9\% primarily use a smartphone to access social media, while 27.9\% use both a smartphone and computer/laptop equally.
    \item \textbf{Distracting Social Media Platform:}
    Instagram was the most distracting social media platform during study time, chosen by 39.3\% of the teenage participants. Other platforms like Facebook (22.6\%) and YouTube (21\%) were also mentioned as major distractions. Only 16.7\% of participants do not have access to a mobile phone or computer during their study time.
    \item \textbf{Mindless Scrolling:}
The majority of teenagers who participated, 45.2\%, spend 1-2 hours mindlessly scrolling on social media daily. A significant 30.6\% of participants spend more than 3 hours engaged in this behavior.

    \item \textbf{Effects on Self-esteem and Well-being:}
When asked whether mindless scrolling has a detrimental effect on their self-esteem or perception of their bodies, 38.7\% of participants said they thought it did. Only 12.9\% of respondents who were teenagers said that mindless scrolling did not influence their self-esteem, while 30.6\% were unsure of its implications. On their overall well-being, 53.2\% believe that mindless scrolling affects them negatively.
    \item \textbf{Triggers for Mindless Scrolling:}
The study identified several content types that trigger mindless scrolling, with inspirational quotes being the most prominent at 53.2\%. Memes and humor (44\%) and celebrity updates (32.3\%) also play significant roles in driving this behavior.
    \item \textbf{Emotions After Mindless Scrolling:}
Emotional responses following mindless scrolling vary among participants. While 37.1\% report feeling neutral, 32.3\% feel annoyed with themselves, and 25.8\% feel guilty after stopping mindless scrolling.
    \item \textbf{Signs of Addiction:}
The survey uncovered signs of social media addiction among the participants. A significant 51.8\% feel restless or troubled if they are prohibited from using social media. A substantial 73.2\% have tried to cut down on social media use without success. Additionally, 57.1\% spend a lot of time thinking about social media or planning how to use it, and 78.5\% feel an increasing urge to use social media more and more. Notably, 71\% even browse social media for study-related purposes but often end up engaged in mindless scrolling.

\end{itemize}

The impact of social media on teenage users' psychological well-being is evident. The exposure to millions of unfamiliar teenagers globally introduces a unique dynamic, fueling social comparison. Adolescents, who are prone to comparing themselves with others, often experience heightened feelings of insecurity and inadequacy. The bright side is that 85.5\% teenage participants show interest in features on social media platforms that encourage mindful and purposeful use. 

\begin{comment}
\subsubsection{Findings from RQ3}
This section focuses on the key outcomes for RQ3 analyzing different demographic factors that influence users’
engagement patterns and responses to infinite scrolling in
social media mentioned in Section \ref{rq3}.
\end{comment}

\begin{itemize}
    \item \textbf{Age Influence Engagement: }Age is a huge factor in content designing and making changes in behavior. Young generations are more prone to use interactive media, engage more easily, are fascinated by tech trendy features, like new and interesting updates, and explore various latest content. In the middle age group, they might find some features complex to use and thus lose interest. A more friendly design is preferred for them. The elderly group finds it hard to use. Their interest level is much lower compared to other classes. Among 67 participants and 8 different age groups, 74.6\% participants were between the age of 18-24 years. As this survey form was shared across different social platforms, it highly indicates that most of the users of social media are between the age of 18-24.  Users from different age groups also exist but not significant. So social media design targets this age audience to engage more with the contents of their interests. Also, 41.8\% of users agreed that they engage more whenever there is content related to their age group.
    \item \textbf{Content preference in Gender Perspective: } There are psychological differences between men and women. There are differences in content preference and thus design preference might vary. Among 67 participants, 62.7\% were male and 37.3\% were female which can conclude that most of the users are male. In an infinite scrolling context, 68.7\% of participants agreed that gender does not influence the response to contents. So, nudging might not include gender differences while designing.
    \item \textbf{Social and Economic Status:} People having different economic and social statuses have different impacts on social media use. Timing of use, digital literacy may lead to making different choices. Also, some have positive impacts, and some have negatives. From the survey, it was concluded that 91\% of participants have mid-level socio-economic status. In Bangladesh, most of the people have a moderate income. So the design patterns might include designs that encourage this portion of users. Showing e-commerce sites from high-range budgets might discourage this group of users as 73.1\% of participants approved that their status influences purchasing decisions on social media. 
\end{itemize}

\subsubsection{Findings from RQ3}
\begin{figure}
     \centering
     \begin{subfigure}[b]{0.45\textwidth}
         \centering
         \includegraphics[width=\textwidth]{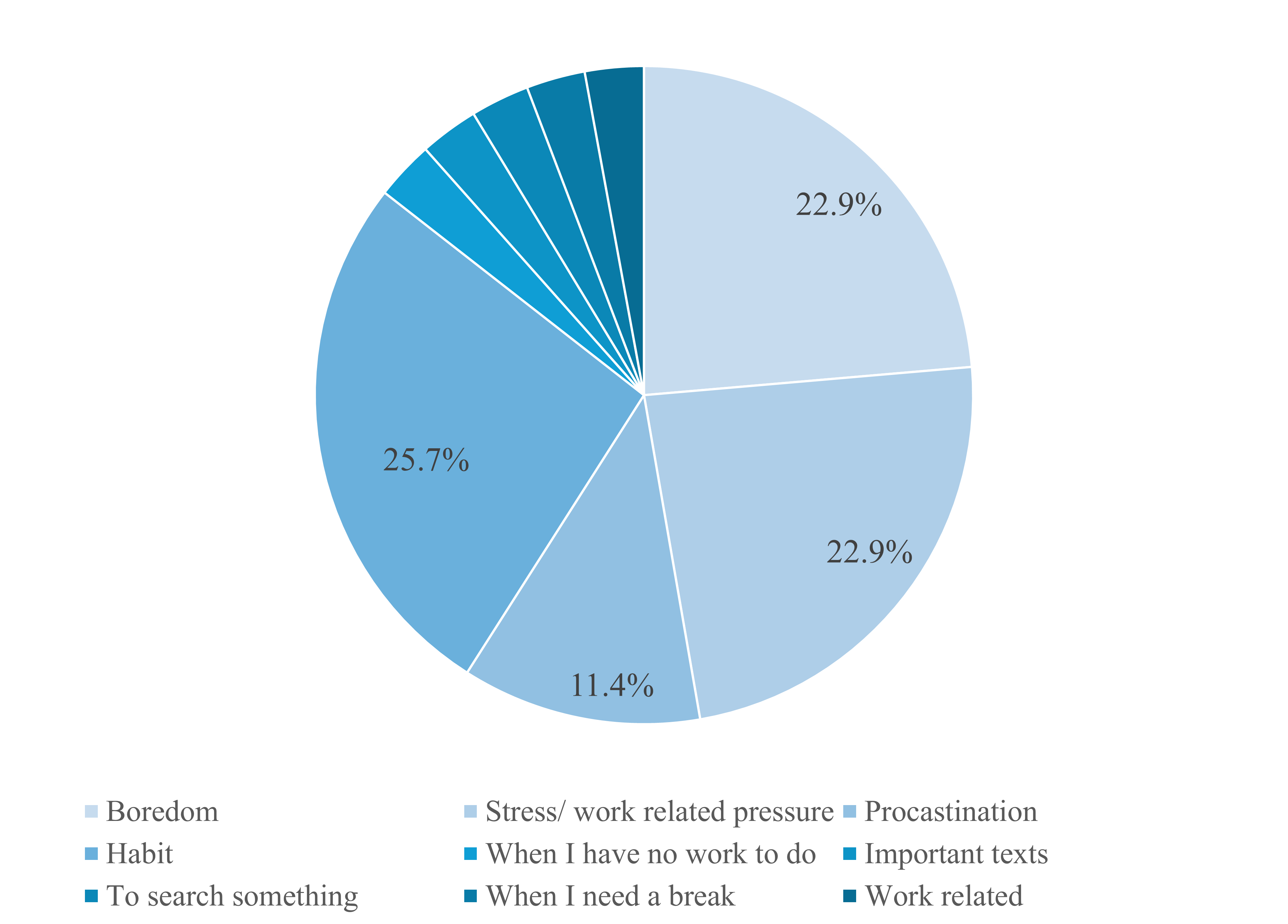}
         \caption{Reasons for prompting work professional to use social media during work hours}
         \label{fig:work-1}
     \end{subfigure}
     \hfill
    
     \vspace{0.5cm}
     \begin{subfigure}[b]{0.4\textwidth}
         \centering
         \includegraphics[width=\textwidth]{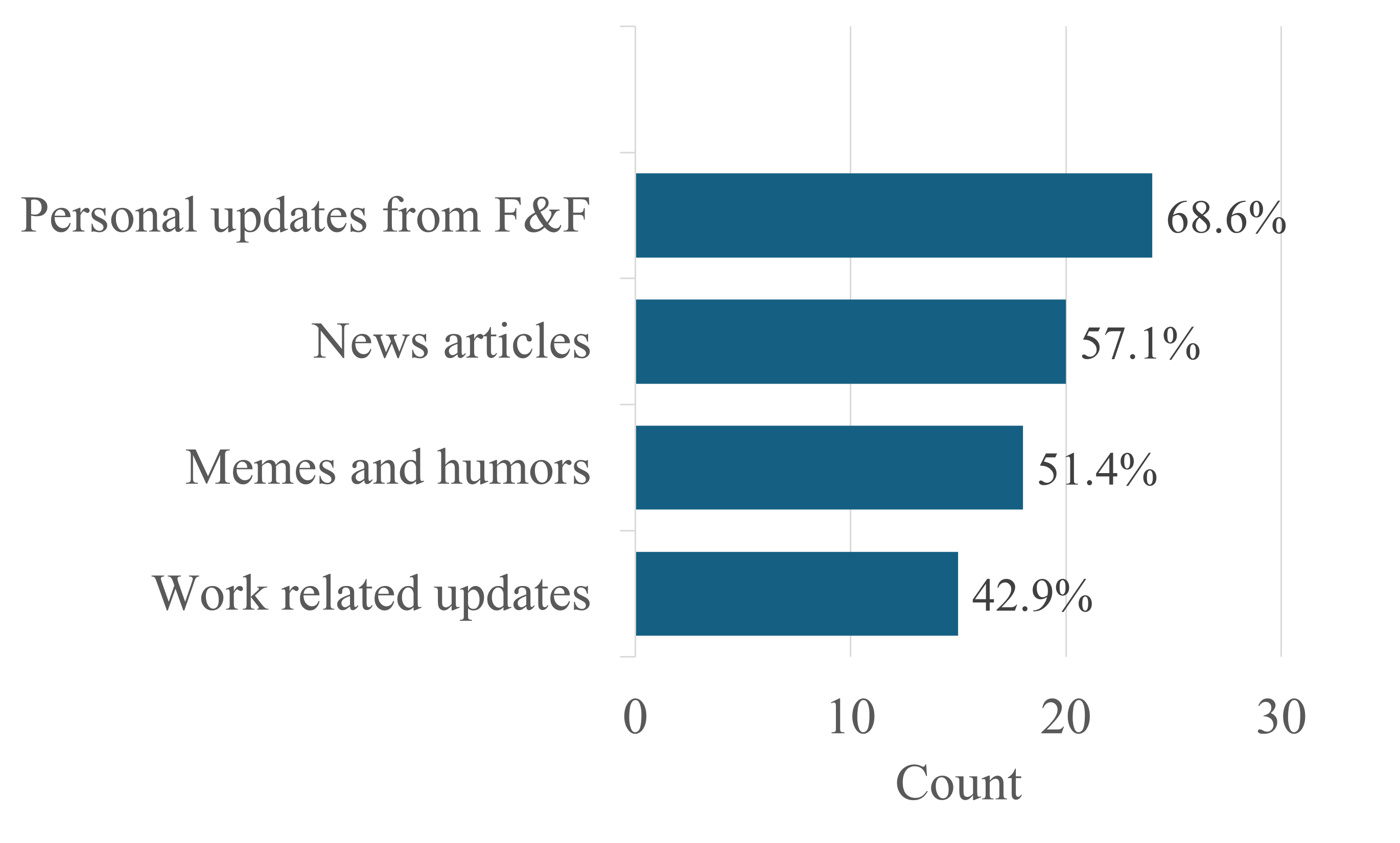}
         \caption{Types of content that typically come across during scrolling at work}
         \label{fig:work-3}
     \end{subfigure}
     \hfill
     \begin{subfigure}[b]{0.4\textwidth}
         \centering
         \includegraphics[width=\textwidth]{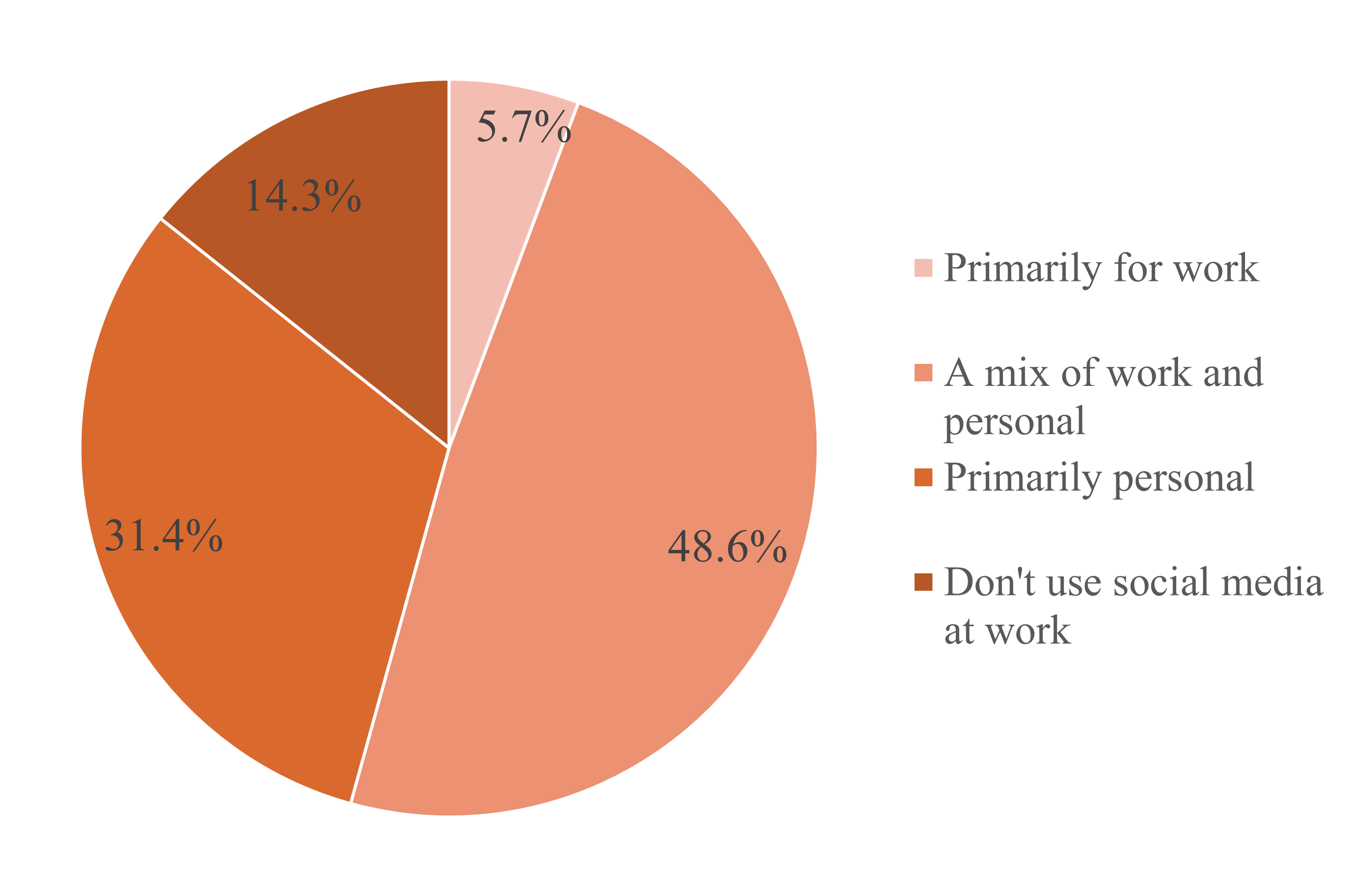}
         \caption{Reasons for using social media during work hours}
         \label{fig:work-5}
     \end{subfigure}
        \caption{Some survey responses on the influence of infinite scrolling of social media on at workplace in Bangladesh}
        \label{fig:work}
\end{figure}

This section focuses on the key outcomes for RQ4 analyzing implementing infinite scrolling on social
media platforms impact productivity, information
consumption habits, and overall work performance of
professional individuals within a workplace context mentioned in Section \ref{rq4}. Figure \ref{fig:work} refers to some significant responses of our study mentioned in Section \ref{rq4}.

\begin{itemize}
    \item \textbf{Participants' Demographics:} The respondents belong to two main age groups: 25-34 (72.2\%), and 35-44 (19.4\%), including both males 58.3\% and females (41.7\%) from Bangladesh.
    \item \textbf{Usage of Social Media: } 
97.1\% of participants use some form of social media platform such as Facebook, LinkedIn, etc.
    \item \textbf{Distracting Social Media Platform:}
Facebook was found to be the most distracting social media network during working hours, with 62.9\% of participants rating it as such. Interestingly, 20\% of participants reported that they do not use social media at all during work.
    \item \textbf{Daily Scrolling Time at Work:}
The majority, 82.8\%, spend up to one hour engaged in scrolling on social media during their work hours.
    
    \item \textbf{Primary Use: } Only 5.7\% use social media during working hours for work-related purposes. Nearly half, 48.6\%, use social media primarily for a mix of work and personal reasons.

    \item \textbf{Triggers for Mindless Scrolling:}
The study identified several content types that trigger mindless scrolling during work. Personal updates from friends and family (68.6\%), news articles (57.1\%), and memes and humor (44\%) were mentioned as the most common triggers.
    \item \textbf{Feelings Guilty/Anxiety After Mindless Scrolling: } More than half of the participants, 51.4\%, reported feeling guilty or anxious when realizing they had spent time scrolling on social media during work hours.

    \item \textbf{Scrolling During Meetings:}
Approximately 22.9\% admitted to occasionally scrolling on social media during work meetings.
        \item \textbf{Impact on Productivity:}
On how mindless scrolling affects productivity, different people had different opinions. Compared to 8.6\% who thought it had a positive influence, 34.3\% who felt it harmed their productivity. 40\% of people felt that scrolling aimlessly did not affect their productivity, which is a sizable percentage.
     \item \textbf{Efforts to Reduce Scrolling:}
When it came to reducing mindless scrolling, 31.4\% of participants reported making successful efforts to cut down on this behavior, while 17.1\% struggled to reduce it.
     
\end{itemize}

When such media consumption hinders our ability to engage in activities that are consistent with our true objectives, a problem arises. We can compare the infinite scrolling feature with the ``rabbit hole'' effect because the social media feeds are enhanced by engaging with several consecutive photographs or short videos, ingesting relatable content, and continuously consuming content. People at the workplace can counteract this by watching only one video, selecting unrelated information for consecutive viewing, or purposely introducing breaks. Using tools like social media timers, visual reminders, and intentionally diverse material might help break the loop. Understanding these processes, in the end, allows people to reclaim control of their content-consuming habits and divert their focus to more productive tasks.

\subsection{Outcomes of Mindful Scrolling}
\label{outcome mindful}
\subsubsection{Findings from RQ4}
\begin{figure}
     \centering
     \begin{subfigure}[b]{0.45\textwidth}
         \centering
         \includegraphics[width=\textwidth]{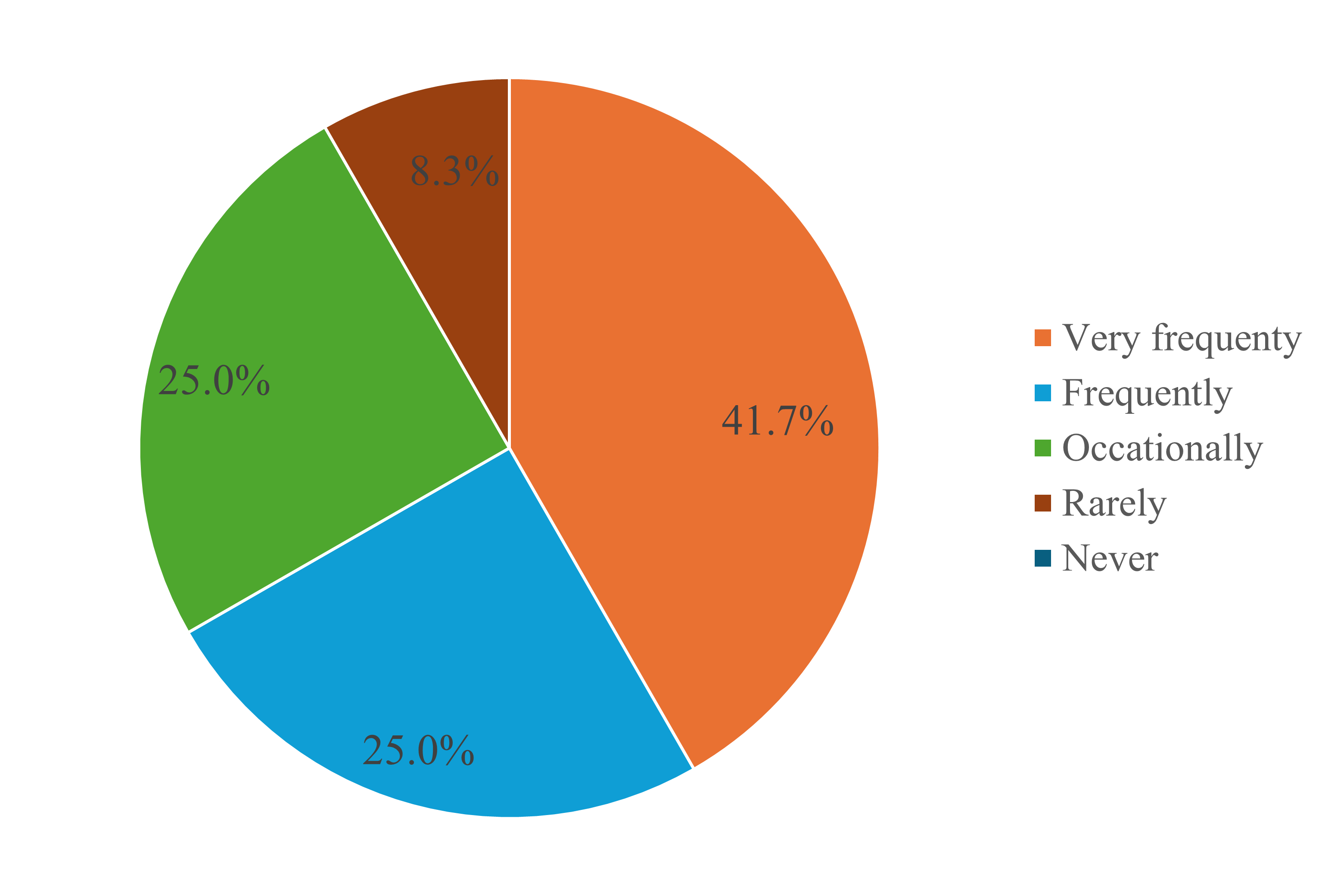}
         \caption{Engagement in mindful scrolling on social media content related to pregnancy or parenting during pregnancy}
         \label{fig:p1}
     \end{subfigure}
     \hfill
     \begin{subfigure}[b]{0.45\textwidth}
         \centering
         \includegraphics[width=\textwidth]{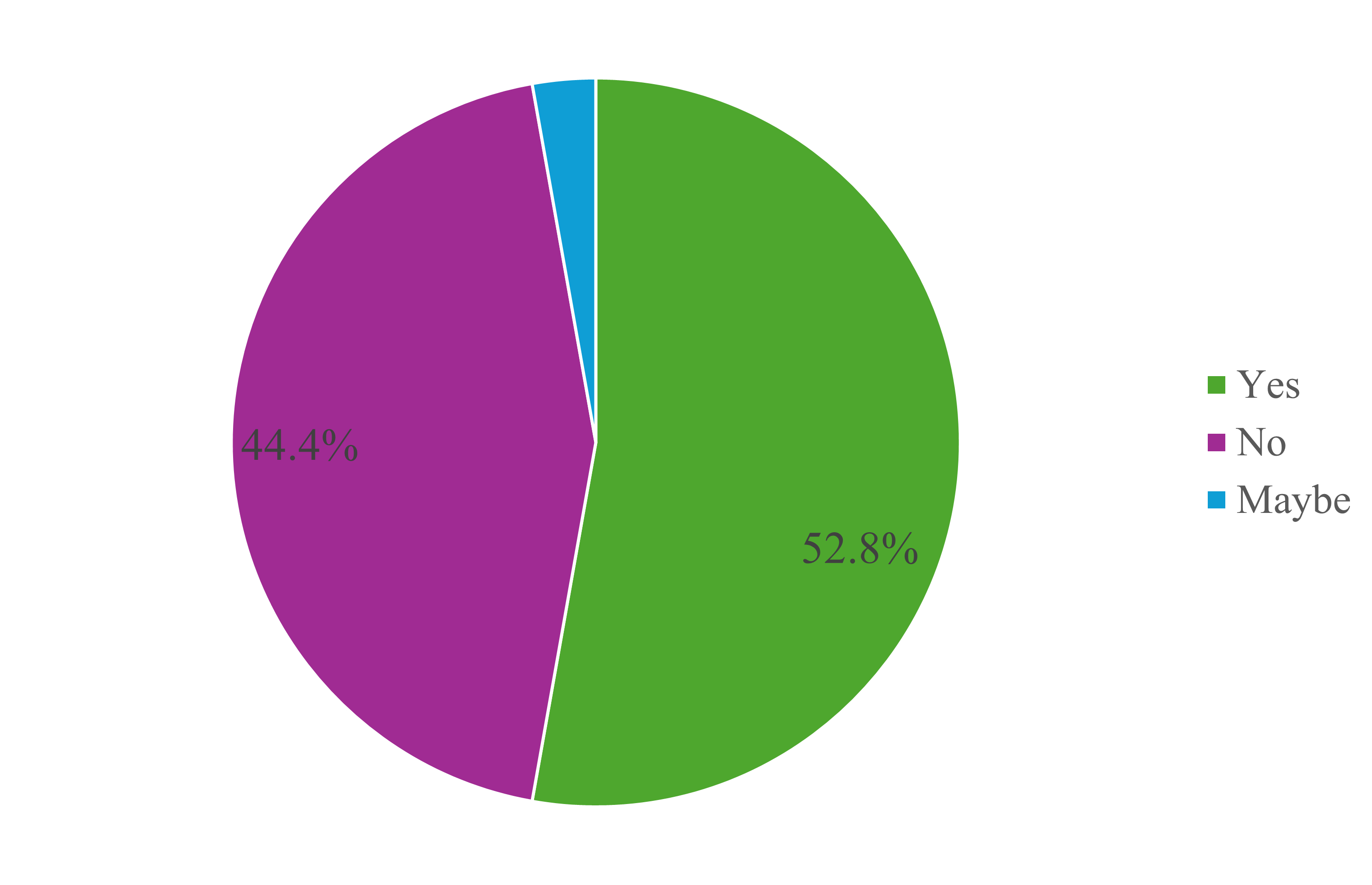}
         \caption{Engaging in mindful scrolling during pregnancy positively influences emotional preparedness for parenting}
         \label{fig:p2}
     \end{subfigure}
     \hfill
     \vspace{0.5cm}
     \begin{subfigure}[b]{0.45\textwidth}
         \centering
         \includegraphics[width=\textwidth]{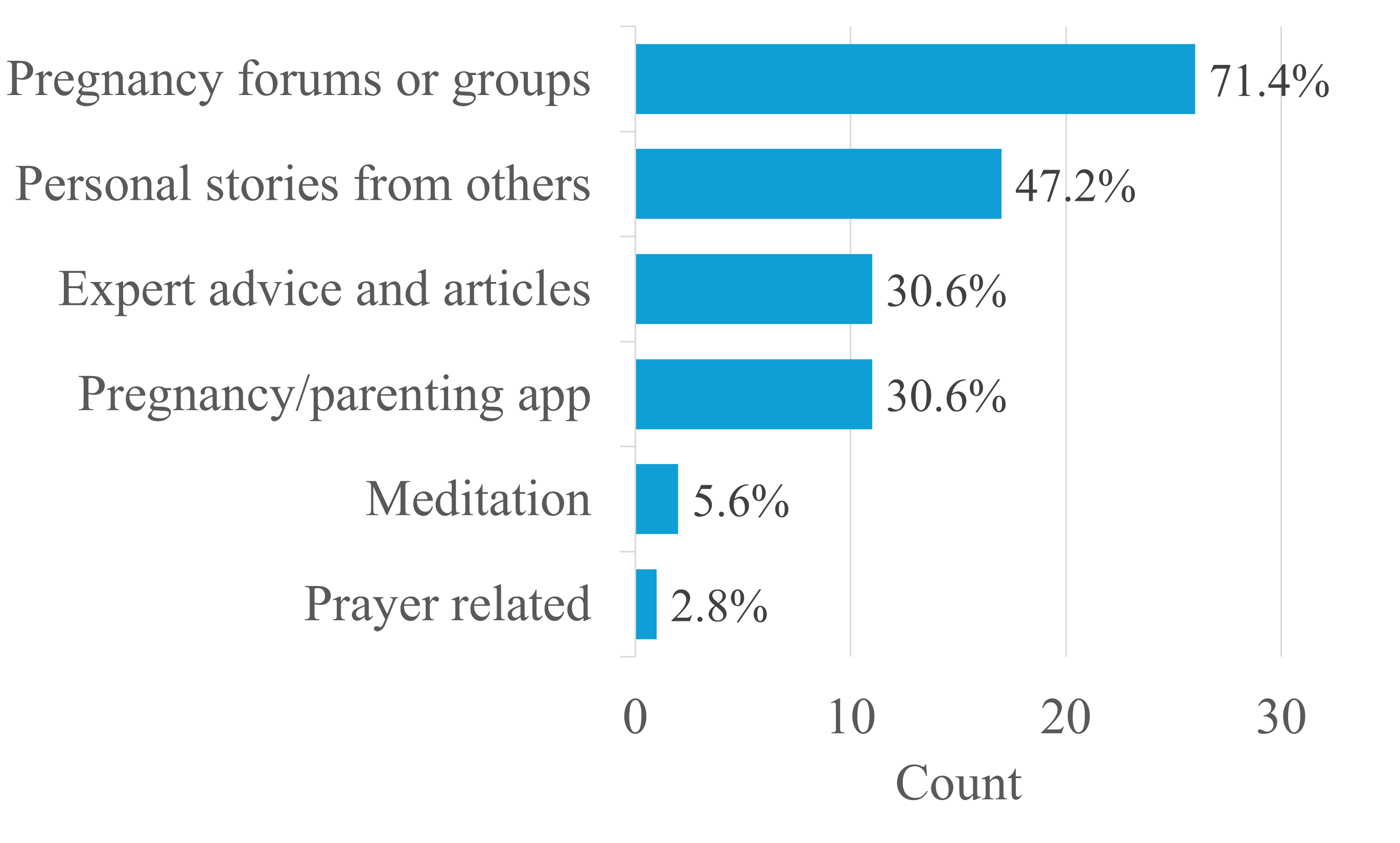}
         \caption{Specific features or types of content on social media found most supportive during pregnancy in terms of emotional preparedness for parenting}
         \label{fig:p3}
     \end{subfigure}
     \hfill
     \begin{subfigure}[b]{0.45\textwidth}
         \centering
         \includegraphics[width=\textwidth]{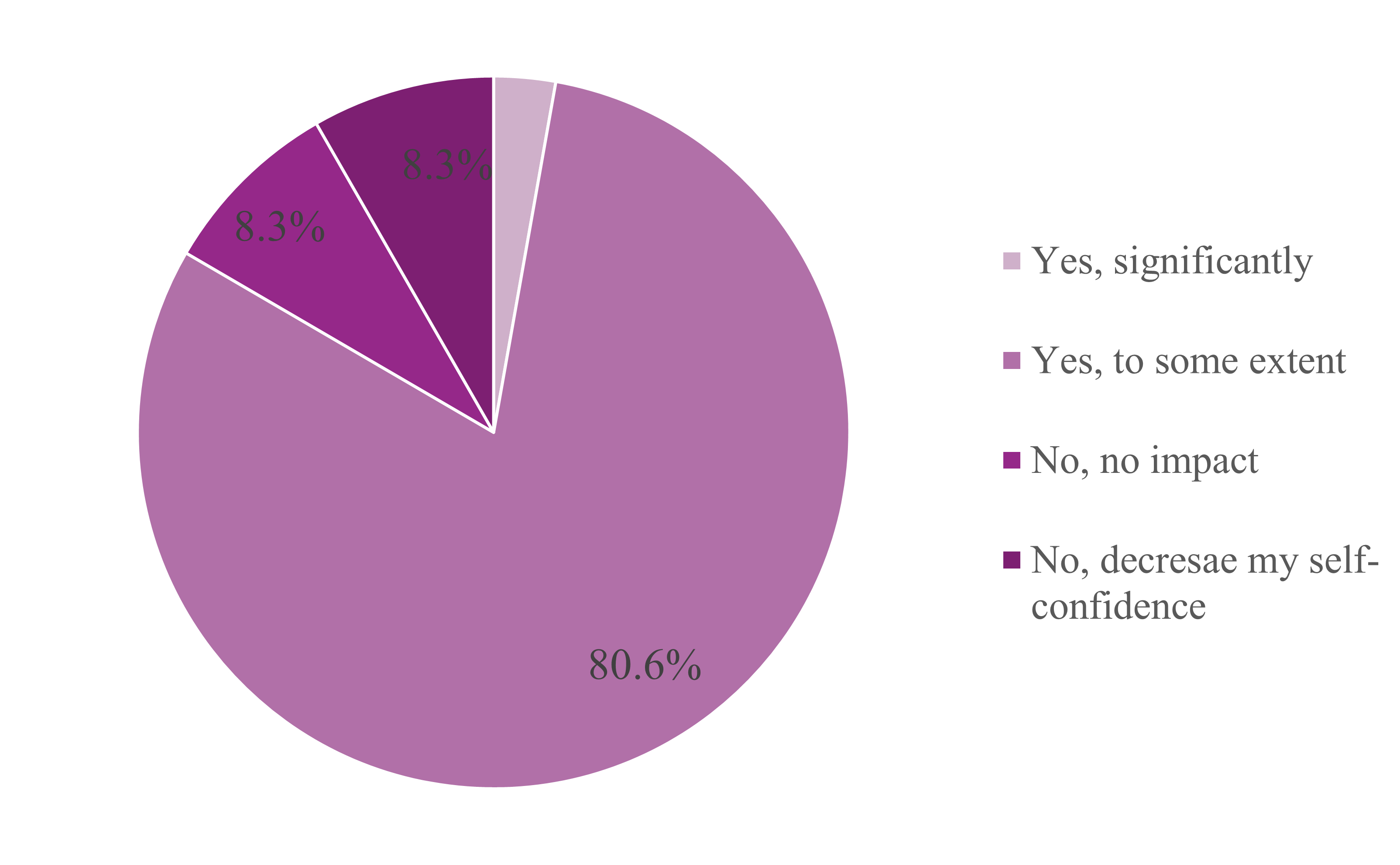}
         \caption{Engaging in mindful scrolling during pregnancy boosts self-confidence in the ability to navigate pregnancy and early parenthood}
         \label{fig:p4}
     \end{subfigure}
        \caption{Some survey responses on the impact of mindful scrolling on emotional preparedness during pregnancy period}
        \label{fig:preg}
\end{figure}

This section focuses on the key outcomes for RQ5 analyzing how mindful scrolling during pregnancy
influences women’s views of their emotional preparedness for
parenting mentioned in Section \ref{rq5}. Some survey responses on the impact of mindful scrolling on emotional preparedness during the pregnancy period are shown in \ref{fig:preg}.

\begin{itemize}
    \item \textbf{Participants' Demographics:} Respondents were mostly between the ages of 25 and 34 (66.6\%) having experience of pregnancy.
    \item \textbf{Scrolling Habit: } 41.7\% of respondents reported they engaged in mindful scrolling habits very frequently while pregnant. 25\% of respondents reported frequently scrolling on social media during pregnancy. This included activities like engaging in online pregnancy forums or groups(72.2\%), reading pregnancy-related content(30.6\%), and following the parenting of other mothers(47.2\%) on social media.
    \item \textbf{Emotional Support and Self-Confidence:} 80.6\% of the respondents believe that frequent access to parenting advice and knowledge can boost a woman's perceived expertise and self-confidence in her parenting abilities as 36.1\% of the respondents reported that emotionally they were very unprepared for motherhood. 52.8\% of respondents are sure that engaging in mindful scrolling during pregnancy positively influences their emotional preparedness for parenting and 44.4\% felt that it may influence them positively.
    \item \textbf{Gathering Knowledge and Community:} Mindful scrolling can enable the opportunity to interact with other expectant parents, share experiences, and acquire a sense of community and emotional support. 41.7\% of the respondents occasionally participated in these works. 80.6\% of respondents reported that the knowledge they gathered through mindful scrolling influenced their decisions regarding parental care, birth plans, or early parenting decisions.
    %\item \textbf{Mixed Emotions: } Constant exposure to a wide range of parenting-related information may cause mixed feelings ranging from enthusiasm and anticipation to nervousness and stress.
\end{itemize}

\subsubsection{Findings from RQ5}
\begin{figure*}
     \centering
     \begin{subfigure}[b]{0.45\textwidth}
         \centering
         \includegraphics[width=\textwidth]{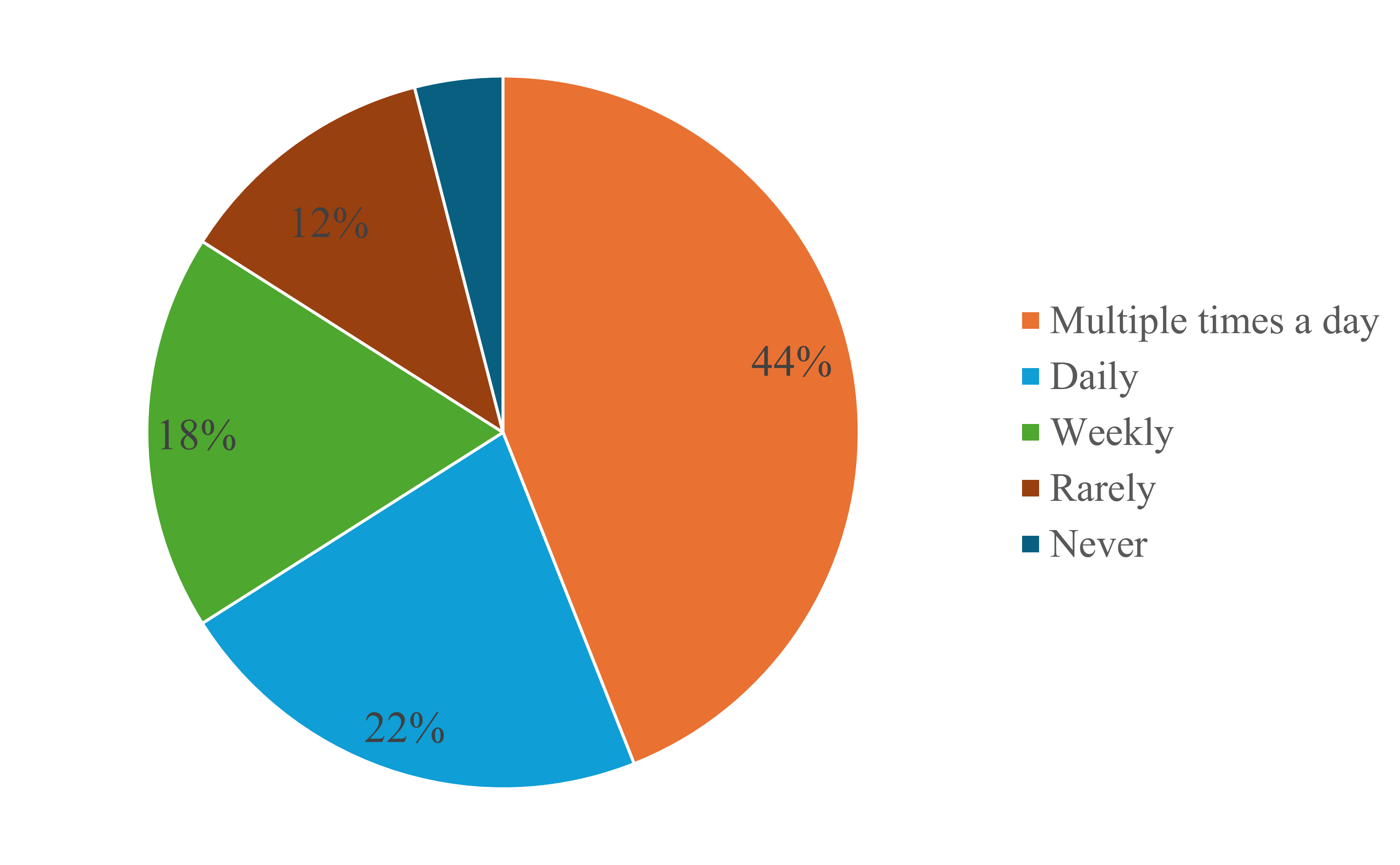}
         \caption{Engagement in mindful scrolling on social media food-trend content}
         \label{fig:d1}
     \end{subfigure}
     \hfill
     \begin{subfigure}[b]{0.45\textwidth}
         \centering
         \includegraphics[width=\textwidth]{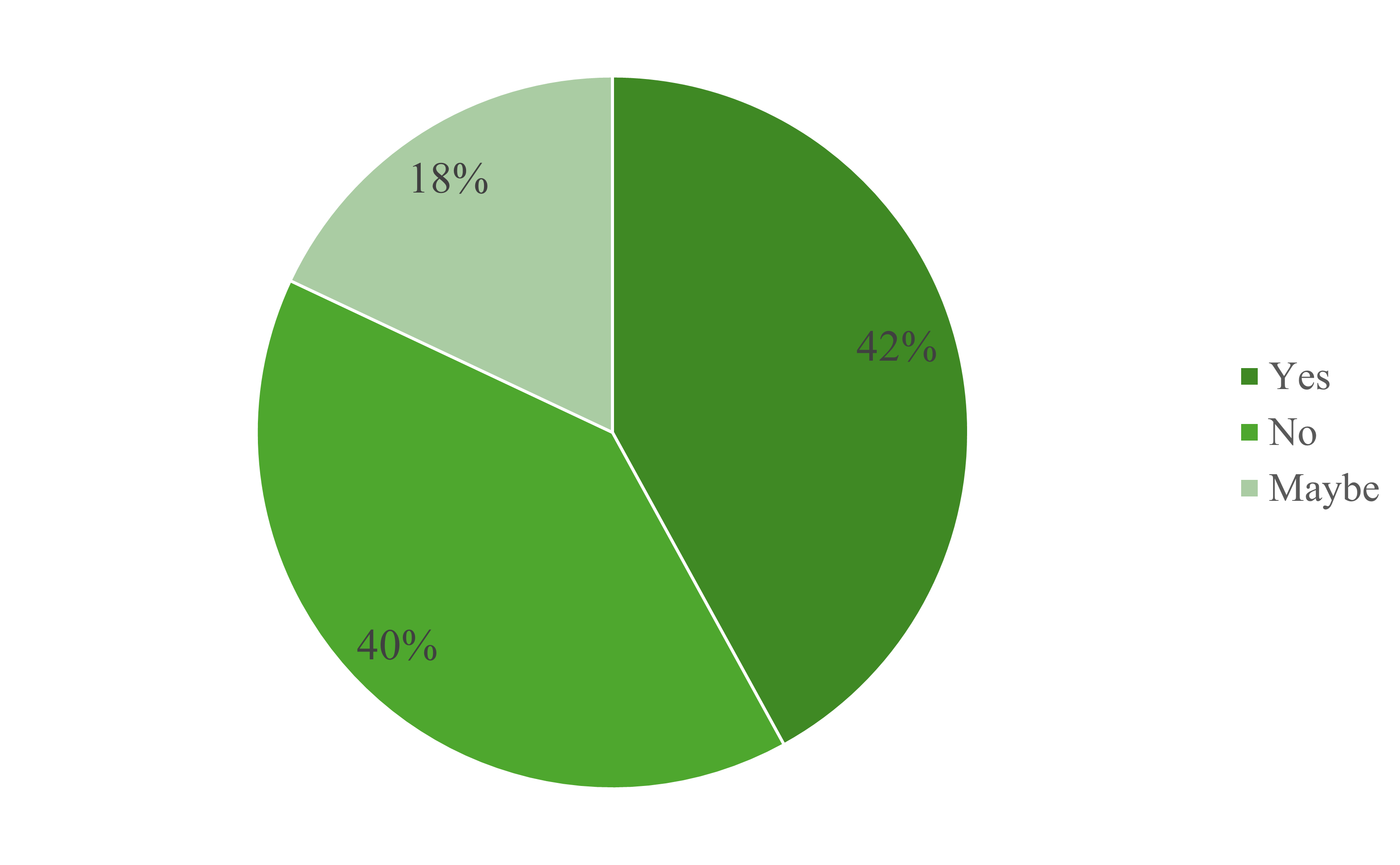}
         \caption{Experienced challenge in dietary goals due to interactions with food trend posts on social media}
         \label{fig:d2}
     \end{subfigure}
     \hfill
     \vspace{0.5cm}
     \begin{subfigure}[b]{0.45\textwidth}
         \centering
         \includegraphics[width=\textwidth]{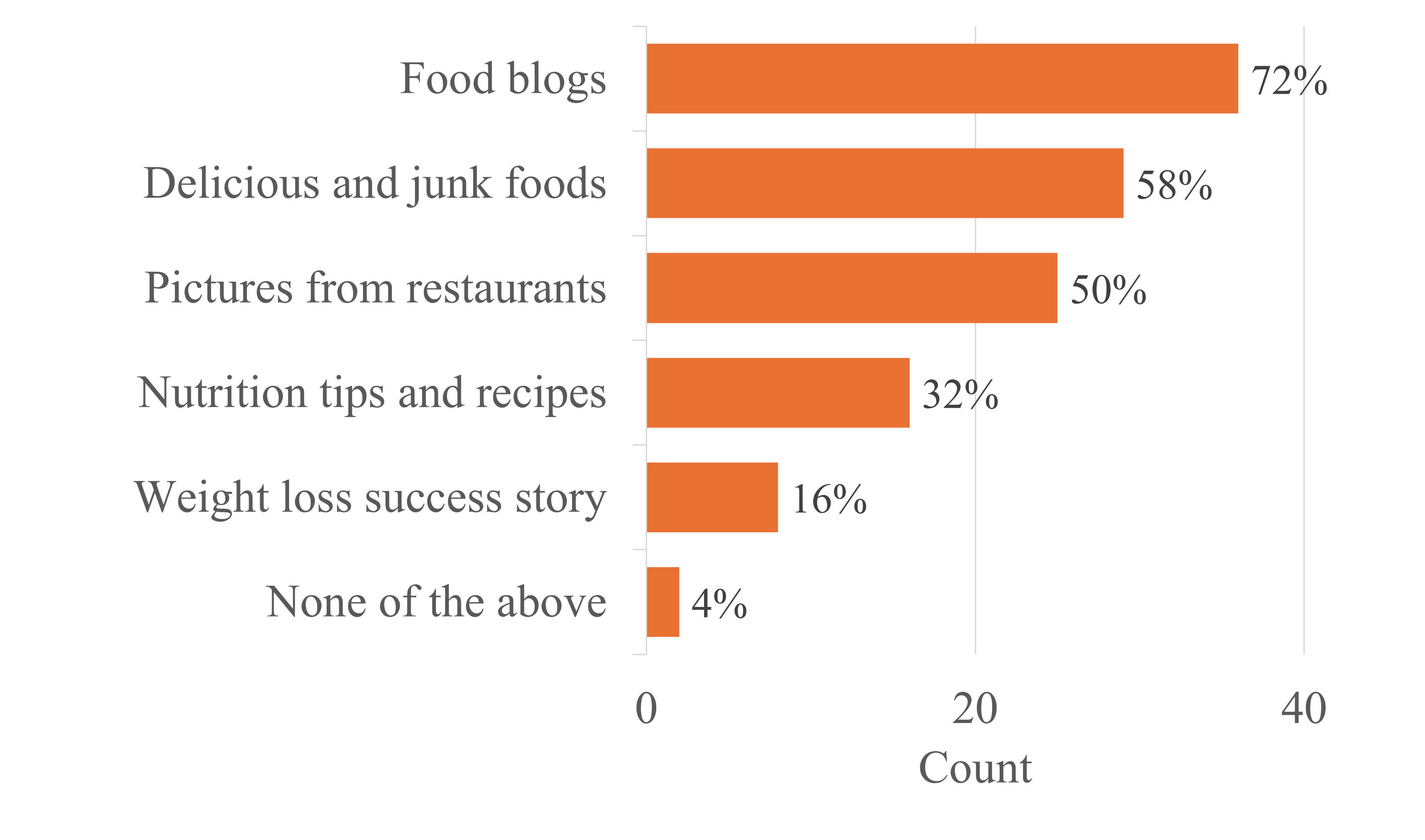}
         \caption{Types of influential content while scrolling on social media }
         \label{fig:d3}
     \end{subfigure}
     \hfill
     \begin{subfigure}[b]{0.45\textwidth}
         \centering
         \includegraphics[width=\textwidth]{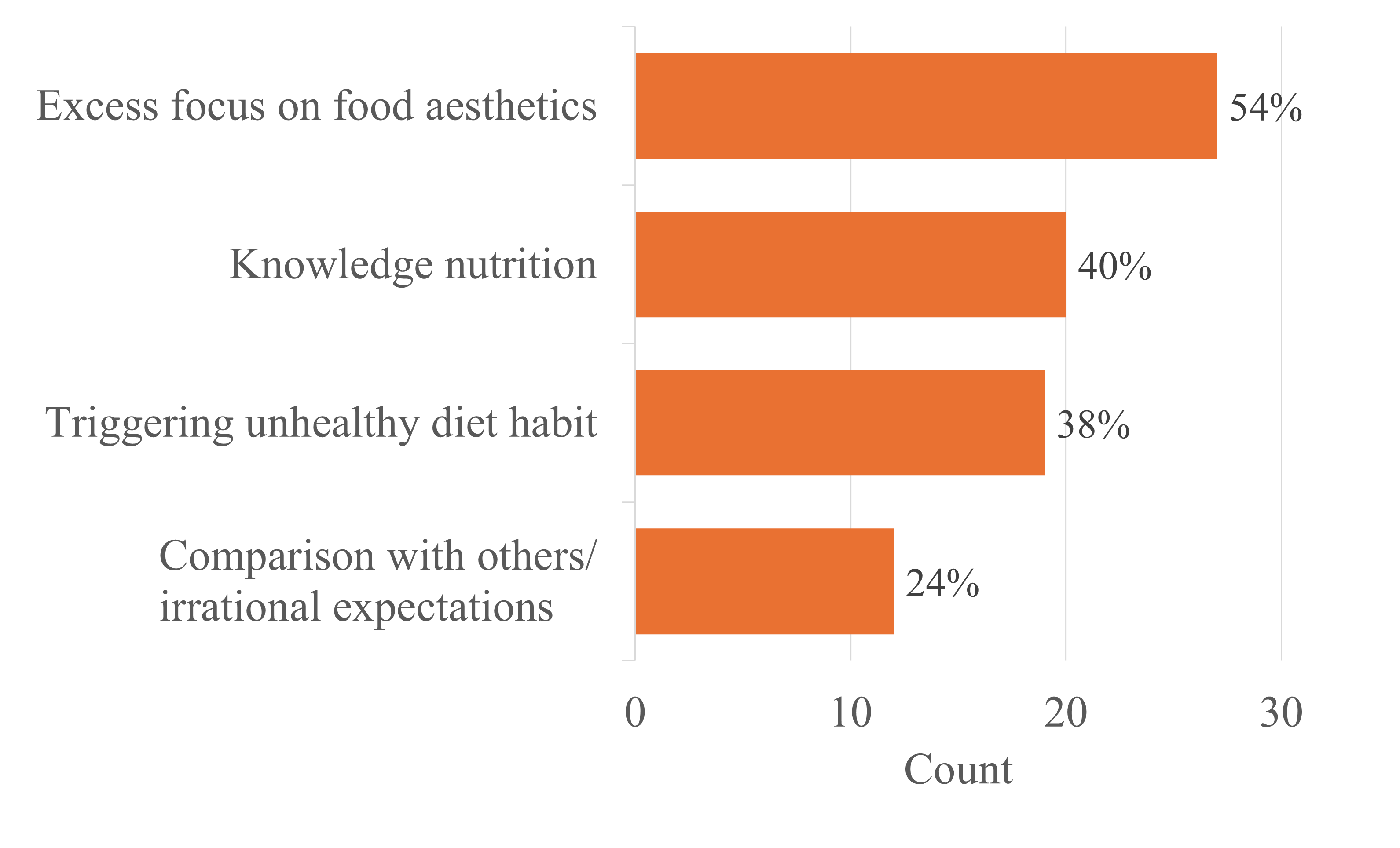}
         \caption{Influence of social media food trend posts}
         \label{fig:d4}
     \end{subfigure}
        \caption{Some survey responses on the impact of mindful scrolling on achieving a dietary goal}
        \label{fig:diet}
\end{figure*}

This section focuses on the key outcomes for RQ5, analyzing the frequent mindful scrolling of social media food
trend posts influence users to achieve a dietary goal mentioned in Section \ref{rq6}. Figure \ref{fig:diet} shows some significant responses to our study.

\begin{itemize}
    \item \textbf{Participants' Demographics:} Respondents were mostly between the ages of 18 and 24 (68\%). 56\% of the respondents are male and the rest(44\%) are female.
    \item \textbf{Social Media Engagement: }A significant proportion of respondents (58\%) reported often engaged in social media for 1 to 3 hours on a daily basis. 44\% of those responded that they often mindfully scrolled through social media food trend postings multiple times in a day. 84\% and 72\% respondents reported that they frequently encounter food-related content on Facebook and YouTube, respectively.
    \item \textbf{Challenges to Achieving Dietary Goals: } 42\% of our respondents are currently working towards achieving a specific dietary goal like weight loss and having balanced nutrition. Due to the interactions with food trend content on social media, 42\% of the respondents agreed that they faced challenges in achieving their dietary goal for encountering different food-trend posts. 18\% of the respondents reported that they may face setbacks for the same reason. 
    \item \textbf{Triggering Unhealthy Diet Habit: }Seeing pictures of delicious food can cause cravings or unsuitable eating patterns in certain respondents(38\%) who engaged in mindful scrolling of food-related posts on social media.
    \item \textbf{Excessive Focus on Aesthetics:} Too much emphasis on food's aesthetic appeal overshadows the significance of balanced nutrition and quantity management reported by 54\% respondents.
    \item \textbf{Comparison and Irrational Expectations: }Constantly watching curated images of current food trends can encourage comparison with others and create irrational expectations for how one's own meals should seem, which leads to feelings of inadequacy reported by 42\% respondents.
    \item \textbf{Knowledge on Proper Nutrition: }Social media can be a good source of information on nutritional advantages, food replacements, and culinary techniques that can help one reach their dietary goals. 40\% of respondents reported that they use social media to gather knowledge on proper nutrition.
\end{itemize}

The intentional design of social media platforms to attract users and influence their decisions has various negative impacts, which we have already analyzed in the earlier research questions'. The rising concern of massive addiction to social media platforms led researchers to investigate the design principles and incorporate a framework in the existing system to promote a sense of healthier interactions among users. Recently, a number of quality researches have been conducted regarding “Mindful scrolling” to combat attention engineering. Therefore, we hypothesized the following research question.

\section{Design Implications based on Our Study}
According to our survey outcomes discussed in Section \ref{outcome}, we offer some suggestions to develop social media features that encourage more positive interaction and limit detrimental impacts.

We have found in our survey, that above 65\% of users know certain design elements are implemented by social sites to keep them engaged in endless scrolling. When we asked them if they were interested in mindful scrolling or not, our results portrayed that above 50\% social media users of all age categories are interested in mindful scrolling for their well-being and digital detoxification. Our survey of teenage participants responded that around 72\% of them tried to control their use of social media but could not succeed. On the other hand, we found only 25\% teenage users acknowledged that they used setting screen time and other simple mindful features to control their endless scrolling, whereas knowing the mindful options most (70\%) of them rarely or never had attempted to use them. The unavailability of user-friendly and systematic implementation of mindful features is one of the reasons for the failure to use them. Therefore, we can conclude that mindful features should be engineered in a clever way to the existing design methods to promote users' involvement in mindful scrolling.\\
\begin{figure} [h]
\centering
\fbox{\includegraphics[width=80mm,scale=1.5]
{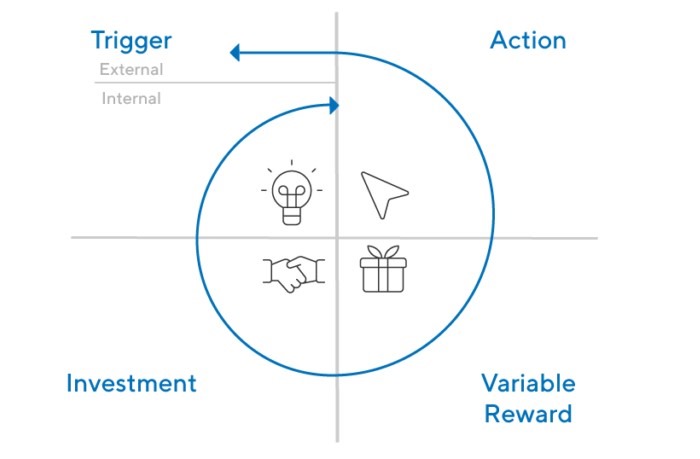}}
\caption{The “hook model” of user engagement in social media} \label{fig:model2}
\end{figure}

  Keeping the existing design principle, Purohit et al.~\cite{purohit2020designing} introduce ``digital nudges'' to promote digital well-being instead of steering people's decisions can be a game changer as it is easy to implement and aware of ethical and privacy matters. Their literature provides different design strategies\cite{purohit2023starving} for a digital detox to lessen social media’s addictive qualities by combating the Hook model's trick, which is graphically represented in Fig.~\ref{fig:model2}. They applied four approaches to break the Hook model’s loop which is also supported by our survey outcomes. Here are the digital detox strategies described as follows: 
\begin{itemize}
    \item \textbf{Dismissing Triggers:}
    Triggers (external and internal) such as notifications, pre-existing routine, and boredom make users periodically checks preferred platform and catch them by the Hook’s loop. In our survey, almost 60\% users agreed that auto-refreshing content triggers them to engage in mindless scrolling. whereas, around 30\% of users blamed it on notifications or alerts. To break the chain, dismissing such triggers is essential. Therefore, hiding nudges can be implemented in order to prevent action from being prompted by the user. However, only hiding nudges is not enough in some scenarios as there are intermediate websites that may lead to access to a social media platform. This scenario is also reflected in our survey where teenage users gave emphasis on \textbf{turning off push notifications} along with other digital nudges to reduce mindless scrolling.

      \item \textbf{Limiting User Action:}
      The personalized content of social media newsfeeds encourages a user to extend his stay on that platform even if he wasn't originally interested, but the customized information could persuade him to read further topics that are connected. Introducing friction can be used to encourage users to engage more intentionally~\cite{purohit2023designing}. For instance, a user may be prompted to indicate if he wants to see suggested movies or postings, which would make mindless clicking more challenging. 

       \item \textbf{Reducing Reward:}
       Social media platforms encourage users to stay longer by giving rewards such as likes, comments, number of visitors views, new content consumption while endless scrolling, etc. Reducing rewards can be achieved by interrupting the actions of users to make them aware of their behavior. Furthermore, a feedback nudge can also be used to provide information about the time spent on an app, potentially discouraging prolonged use.  

    \item \textbf{Reduce Investment:}
    To break the loop of Hook’s model investments of users should be diminished. Social media users continue visiting the platform as they invest in it by following friends and pages. This strategy is also supported by our survey participants as they responded that most of them (67\%) unfollow their friends and pages for a digital detox at the present time. Thereby, giving an automated unfollow option in place of the tedious unfollowing procedure that is now required on most platforms. 
\end{itemize}

Moreover, we can recommend some features for social media encouraging positive interactions and lessen harmful impact.
\begin{itemize}
    \item \textbf{Time Limit and Session Goal Implementation:} Social media platforms may include tools that allow users to establish session-based time limitations for their social media activity. Moreover, the features that allow users to establish particular goals may be implemented before beginning their session, such as discovering new recipes, staying up to date on current events, or learning a new skill. This tool can help users focus on reaching their goals rather than mindless scrolling.
    \item \textbf{Filtered Recommendation:} The auto recommendation of feeds can be filtered for some specific period. For example, if someone is struggling to achieve a dietary goal, the recommendation system should filter the contents which trigger the unhealthy food habits. 
    \item \textbf{Thoughtful Prompts: }Social media platforms can encourage visitors to consider their feelings while they are scrolling by introducing prompts on a regular basis. According to the feelings provided by the visitors, the feeds can change the topics for balancing mental and emotional well-being.
\end{itemize}

\section{Conclusion}
This study demonstrates how user behavior is greatly influenced by the design of social media platforms, including features like algorithmic personalized service, auto-refreshing feeds, and infinite scrolling.  According to our findings, these design techniques frequently result in mindless scrolling, which exacerbates emotional issues, including stress, anxiety, exhaustion, and guilt, especially among Bangladeshi teens and working adults.  However, we additionally found that scrolling mindfully can have beneficial outcomes, such as improved dietary habits, motivation, and exposure to educational materials.

Unlike previous studies that focused on digital addiction or problematic use of smartphones, our study is the first to examine specific scrolling behaviors (mindless vs. mindful) in a variety of demographic groups, such as pregnant women, teenagers, and professionals working in the Bangladeshi context.  Through the integration of insights from the literature on behavioral design with the viewpoints of local users, this paper contributes to the expanding research on digital well-being and attention engineering.

In conclusion, to improve user pleasure and reduce psychological damage, we suggest using design elements that promote mindful involvement, such as instruction nudges, time-awareness indications, and content restriction.  Future social media companies need to intentionally build their systems to promote healthy user habits to balance engagement and well-being.
\section*{Declaration}
\subsection*{Clinical Trial Number}
Not Applicable

\subsection*{Consent to Publish declaration}
Not Applicable

\subsection*{Competing Interests}
Not Applicable

\subsection*{Funding Information}
Not Applicable

\subsection*{Data Availability Statement}
The survey responses gathered and investigated during the current study are not publicly available due to privacy and ethical restrictions protecting participant confidentiality.

\subsection*{Research Involving Human and/or Animals}
Not Applicable

\subsection*{Informed Consent}
This study did not involve any clinical interventions and was conducted using survey-based methods. The research was approved by the Institutional Review Board of Bangladesh University of Engineering and Technology (BUET). All procedures were performed in accordance with the ethical standards of the 1964 Helsinki Declaration and its subsequent amendments. All participants provided informed consent prior to participation. For participants under 16 years of age, consent was obtained from their legally authorized representatives. Participation was entirely voluntary, and all collected data were anonymized to ensure confidentiality.

\subsection*{Author Contributions}
Sanzana Karim Lora: Writing-review and editing, Writing-Original Draft, Data Collection, Visualization, Methodology, Investigation, Formal Analysis, Conceptualization, Supervision; Sadia Afrin Purba: Data Collection, Visualization, Conceptualization; Bushra Hossain: Writing-review and Editing, Data Collection, Methodology, Visualization, Conceptualization; Tanjina Oriana: Reviewing-Original Draft, Methodology, Validation, Formal Analysis; Ashek Seum: Data Analysis; Sadia Sharmin: Reviewing, Supervision

\bibliographystyle{bst/sn-aps}
\bibliography{sn-bibliography}% common bib file

\end{document}